\documentclass[11pt]{article}

\pdfoutput=1

\usepackage{abstract}

\usepackage{helvet}
\usepackage{rotating}

\usepackage{xcolor}

\usepackage{pgfplots}
\usepackage{tikz}
\usetikzlibrary{external}

\newcommand{\reminder}[1]
{{\hspace{0pt}\marginpar[$\blacktriangleright$]{\mbox{$\blacktriangleleft  \blacktriangleleft \blacktriangleleft$}}}}

\usepackage{amsthm}
\usepackage{amsmath}
\usepackage{amssymb}
\usepackage{ulem}
\usepackage{helvet}
\usepackage{rotating}
\usepackage{url}
\usepackage{breakurl}
\usepackage{caption}
\usepackage{subcaption}
\usepackage{cite}
\usepackage{color}
\usepackage{setspace}
\usepackage{booktabs}

\usepackage{graphicx}
\usepackage{sidecap}
\usepackage[update,prepend]{epstopdf}
\usepackage{subfig}
\usepackage{tabularx}

\usepackage{pgf}
\usetikzlibrary{positioning,arrows,automata,3d,mindmap,trees,decorations.markings,shapes,plotmarks}

\usepackage{stmaryrd}
\usepackage[mathscr]{eucal}

\newtheorem{theorem}{Theorem}

\newcommand{\matr}[1]{\mathbf{#1}}
\newcommand{\tensor}[1]{\boldsymbol{\mathscr{#1}}}

\makeatletter
\renewcommand{\@biblabel}[1]{\quad#1.}
\makeatother

\date{\begin{center} First version: November 2013. This version: \today \end{center}}

\begin{document}

\title{\textsc{Opinion Dynamics and Price Formation: a Nonlinear Network Model} \\ \bigskip \small \textbf{Draft - please do not quote or cite without permission}}
 


\maketitle

\author{\begin{center}Marco D'Errico\footnote{Corresponding author: m.derrico2@campus.unimib.it} \quad Silvana Stefani \quad Giovanni Zambruno
\\Department of Statistics and Quantitative Methods, University of Milano - Bicocca\\ \bigskip Gulnur Muradoglu \\ School of Business and Management, Queen Mary University of London \end{center}}

\vspace{1cm}

\begin{abstract}
Opinions and beliefs determine the evolution of social systems. This is of particular interest in finance, as the increasing complexity of financial systems is  coupled with information overload. Opinion formation, therefore, is not always the result of optimal information processing. On the contrary, agents are \textit{boundedly rational} and naturally tend to observe and imitate others in order to gain further insights. Hence, a certain degree of interaction, which can be envisioned as a network, occurs within the system. Opinions, the interaction network and prices in financial markets are then heavily intertwined and influence one another.
We build on previous contributions on adaptive systems, where agents have hetereogenous beliefs, and introduce a dynamic confidence network that captures the interaction and shapes the opinion patterns. The analytical framework we adopt for modeling the interaction is rooted in the opinion dynamics problem. This will allow us to introduce a nonlinear model where the confidence network, opinion dynamics and price formation coevolve in time. A key aspect of the model is the classification of agents according to their topological role in the network, therefore showing that  \textit{topology matters} in determining how of opinions and prices will coevolve. 
We illustrate the dynamics via simulations, discussing the stylized facts in finance that the model is able to capture. Last, we propose an empirical validation and calibration scheme that makes use of social network data.
\end{abstract}

\vspace{0.5cm}

\begin{center} \textbf{Keywords} \\ \smallskip opinion dynamics, network, price formation \end{center}

\normalsize

\pagebreak

\tableofcontents

\pagebreak

\section{Introduction}

\bigskip

\begin{quotation}
\noindent \textit{If men define situations as real, they are real in their consequences}.\\ W. I. Thomas and D. S. Thomas (1928, p. 572)
\end{quotation}

\bigskip

Social systems are endogenously driven by how agents perceive, think of and, ultimately, model the system itself. Therefore, the interpretation of a particular situation yielding to a subsequent action often affects the situation. However, the interpretation is almost never \textit{objective} and carries a \textit{certain degree of subjectivity} stemming from individual choices and behaviours. This naturally reflects on the aggregate behaviour of the system. Whether accurate or not, individual views, opinions and beliefs  have \textit{consequences} with a \textit{real} and tangible impact on the system. The quote reported above is also known as \textit{Thomas theorem}.\footnote{For further details, see Robert K. Merton's 1995 historical account on the Thomas theorem where the author explains why the statement became known as a \textit{theorem}.} This ``theorem''  refers to a broad situation where opinions determine reality and, in turn, reality shapes opinions.

This problem can be envisioned within the framework of \text{rationality}, arguably one the most debated concepts in both social and economic theory. A ``rational'' -- in any acceptation of the word -- agent would never act in such a way that a potential undesired consequence could originate from the wrong definition of the situation. This idea, in the light of the current financial crisis, is even more challenged within the standard rationality paradigm in finance, i.e. that investors make optimal use of all available information.

The traditional financial paradigm is built upon the assumption of \textit{rationality} of agents and markets that are perfectly efficient (Fama, 1970). In the Efficient Market Hypothesis (EMH), asset prices at any time reflect all the information available to the agents in a financial system. However, recent studies have shown that investors' behaviour often deviates from the rationality implied by the Efficient Market Hypothesis.  The idea that investors' psychology plays a fundamental role in determining price dynamics has led to a new field called \textit{behavioural finance}, where the behaviour of agents in a financial system departs from the assumption of rationality.

Behavioral aspects can explain a wide range of financial phenomena that cannot be interpreted within the EMH framework (e.g.  momentum trading, trend extrapolation, noise trading, overreaction,  overshooting, contrarian strategies). Some useful references include DeBondt and Thaler (1985),  Hong and Stein (1999, 2003), and review papers such as Hirschleifer (2001), and Barberis and Thaler (2003). Behavioural finance is currently expanding into different directions (DeBondt \textit{et al.}, 2008).

The debate is enriched by the fact that financial systems are witnessing the emergence of very complex structures.
Complexity definitely plays a key role here: the crisis has unequivocally shown that -- within the complex financial environment -- it is unrealistic to assume that, even in presence of full information, agents will have the cognitive capabilities to process and make optimal decisions from the available informative set.

There is, in fact, a strong consensus that financial systems show an increasing degree of complexity on at least two levels. The first level relates to the rapid expansion in the number of financial products traded. The intrinsic complexity embedded in the financial evaluation of such products has remarkable consequences in terms of systemic risk (Caccioli \textit{et al}. 2009; Brock \& Hommes, 2008). The second level pertains to the complex network structure of financial markets (Haldane and May, 2011; Battiston \textit{et al.} 2013).

On both levels, complexity in financial markets implicates that  a huge amount of information needs to be gathered, analyzed and understood before even attempting to make any decision, let alone an \textit{optimal} one. In this increasingly complex  environment, hence, the view  that investors are fully rational implies that they must possess unrealistic cognitive skills. 

The key aspect here is precisely linked to the limited cognitive capabilities of human beings. Prompted by such limitations, Herbert Simon (1955) introduced the idea that agents have ``limited knowledge and ability''. This approach led to the introduction of the concept of \textit{bounded rationality}, which entails the viewpoint that agents would rather adopt a ``satisficing''  strategy as opposed to a fully optimal decision-making strategy. Simon refers to environmental conditions leading to this particular behaviour. More specifically, Bawden and Robinson (2008) refer to \textit{information overload} as a core characteristic yielding to boundedly rational behaviour. Agnew and Szykman (2005) have explored this concept in the context of behavioural finance, stressing that informative overload prevents investors from formulating and deciding upon the correct strategy to pursue and might result in more simplistic investing strategies.

The complexity conundrum implies that some investors might look and imitate the behaviour of those investors who are believed to be better informed, sometimes ignoring fundamentals or private information. Devenow and  Welch (1996) refer to such situation as \textit{rational} herding and, interestingly within the context we intend to model, stress the importance of \textit{informational (cascade) learning}. Crucial for our analysis, these works show that investors might consider ``pooling'' other investors' opinions for a number of reasons, including the fact that they are unsure about the information they possess, for strategic purposes or even because they show a particular aversion for large losses (Kahneman \& Tversky, 1984).

Sobel (2000) wonders whether imitation can be attributed to a rational choice and  emphasises the separation between ``individual'' and ``social'' learning. In this light, agents are ``informationally linked, so that the actions and payoffs of one agent provide information about the state of the world'' to other agents. Sobel also raises the issue of whether individual na\"{\i}veness of agents in large groups translates into the system being ``smart'' in the aggregate.

This is the object of the study of a seminal paper by Galton (1907) about the so--called ``wisdom of the crowd'' (Surowiecki, 2005) enigma. Galton found that the average guess, among those expressed by a group of people upon the butchered weight of an ox, was actually an extremely accurate way to gauge the actual weight.

Are then aggregate predictions always so \textit{accurate}? If we observe  the aggregate behaviour in certain critical situations -- and, in financial system, also in non-critical ones -- one should rather speak of ``madness of crowds'' (cit., Mackay, 1841).

Golub and Jackson (2010) tackled the problem of na\"{\i}ve learning in connection with the ``wisdom of crowds'' in a social network framework. In particular, they determine conditions when beliefs converge to a true value and what obstacles might hamper the convergence process.  Lorenz \textit{et al.} (2011) propose an empirical validation of this problem.

This work and, later,  the contributions of Hegselmann and Krause (2005) and Lorenz (2005a, 2005b, 2006) provide with detailed analytical conditions for the convergence towards the consolidation or the fragmentation of opinion patters. The opinion dynamics  modeling framework used in these works is routed within a stream of literature where the key motivation is the search of a common opinion, or \textit{consensus}. Perhaps, the most well known formulation is the one proposed by DeGroot (1974). However, several authors had previously tackled this problem (notably French, 1956; Harary, 1959). Interestingly for our problem, the basic idea beneath DeGroot's approach is that agents ``pool'' the opinions of neighbours and revise their own opinion by a simple arithmetic average of these opinions. 

This modelling scheme constitutes the basis of our work and motivates Golub and Jackson's approach within a social network framework, as it implies a \textit{pairwise} interaction among agents. The interaction is represented by the \textit{confidence} weights agents place onto one another.

Later contributions (namely the already mentioned Hegselmann and Krause, 2005; Lorenz, 2007) extend this approach, by introducing the idea that the confidence network depends on the opinions and,  in turn, opinions are updated according to the network. Nonlinearity arises precisely in that opinions influence the confidence network and vice versa. This will be the basis of the nonlinear model we propose. The scheme proposed Hegselmann and Krause (2005) is based on an \textit{update rule} according to which, at each time, every agent select a subset of agents whose opinions are the closest, within a specified threshold.  

For this reason, this opinion dynamics modelling approach is referred to as a \textit{bounded confidence} scheme. We can therefore state that bounded confidence is one of the many \textit{degrees of freedom} (Barberis and Thaler, 2003, p. 64) in behavioural finance that can be associated to \textit{bounded rationality}. The conceptual link between the change in confidencewith \textit{bounded rationality} is then made explicit in this class of models. 

The basic argument here is that the bounded rationality of agents in the marketplace naturally leads to a certain level of interaction among agents. The mutual interaction between agents is a key aspect within this context and will be the main subject of our study. 

Several authors have investigated this problem under different perspectives and modelling approaches. Lux and Marchesi (1998) show that scaling power law properties and time-dependent volatily may depend on this interaction. In view of this problem Cont and Bouchaud (2000) model \textit{herd} and imitative behaviour with a random interaction structure among clusters of agents, finding a potential explanation for heavy tails in the distribution of returns. 

Analogously to the bounded confidence model, agents' interaction can be the result of a specific \textit{selection}. As such, a model featuring an \textit{adaptive} financial system, is more suitable at explaining these behavioural limitations (Hommes and Wagener, 2008). In particular, an emergent stream of literature deals with models where heterogenous agents interact and determine asset price dynamics. Among the several contributions in this field, those of Brock and Hommes (1998), Lux (1998), Gaunersdorfer and Hommes (2007) and Chiarella \textit{et al.} (2007) are of particular interest for our work. 

The introduction of \textit{heterogeneous} agents finds its root in overcoming the typical limitations embedded in models with representative agents: as a matter of fact, in the real world, agents naturally have heterogeneous beliefs about asset dynamics. In this context, a logical separation between two categories of agents arises: \textit{fundamentalists} (i.e. agents who believe that prices will eventually converge towards a fundamental value, simply given by the discounted sum of future dividends) and \textit{chartists}. Under the EMH, the latter category would be \textit{irrational} as they would simply lose money in the long run. However, there is evidence that these agents might be able to obtain higher-than-average returns. 

Consistently with the approach we will follow, one of the basic tenets of this class of models is that agents want to exploit temporary deviations from a fundamental price (based on rational expectations) and this might lead agents to adopt short--period strategies that depart from rationality (see Brock and Hommes, 1998 for a thorough discussion).

These \textit{Agent Based Models} (ABMs)  often feature an adaptive dynamics (adaptive beliefs) based on the \textit{success} of previous strategies (either in simple terms of price forecast or in terms of accumulated profits) where the \textit{interaction} in these models stems from the agents' capability of observing other agents' strategies and, by switching strategy, imitating the most successful ones. However, the number of \textit{agent types} presented in these model is usually limited and the interaction occurs via price realizations. 

We propose to overcome these limitations by modelling agents' interaction via a dynamic network approach, associating the network process to the price dynamics. The pricing equation is based on opinions (Section \ref{sec:model}) formed upon a certain rule, therefore introducing the nonlinear network opinion dynamics approach within this class of models.

In our model, agents interact along three intertwined dimensions: i) a network, based on the level of \textit{confidence} that agents place onto one another, ii) a more classical interaction via the prices, and iii) comparing their opinions.

An important caveat should be made clear. In the model we propose, we are not strictly interested in whether and how the system reaches a consensus of opinions, as opinions in our model can vary significantly in time, although this represents surely an interesting case and will be explored in the case of reversion towards a fundamental financial value. We are rather interested in finding a way to characterize agents according to their role in the nonlinear opinion/price/network process, according to their \textit{topological role} within the network. 

The model captures also situations when changes in the topology are coupled with significant shifts in opinions. This is particularly important within the increasing adoption of high frequency trading. This aspect will be further explored in the following motivating example, where we draw a parallel with the aforementioned Thomas theorem.

\smallskip

\subsubsection*{A motivating example}

\begin{figure}[!htbp]
\centering
\includegraphics[scale=0.7]{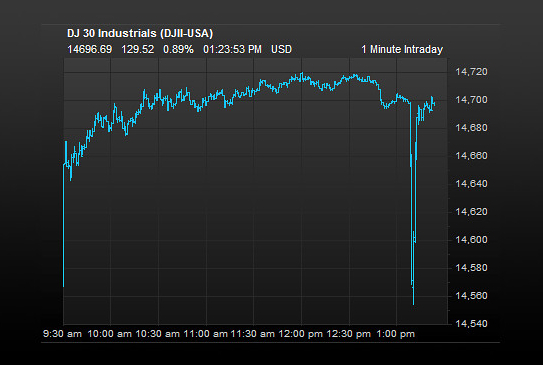}
\caption{April 23 DJ 30 Industrial crash -- 1 minute intraday.}
\label{fig:djcrash}
\end{figure}

On April 23, 2013, at 1:07 PM the US stock market crashed by about $1\%$ in a couple of minutes, recovering soon afterwards in about the same time.  Fig. \ref{fig:djcrash} shows a snapshot of the DJ 30 industrial index one minute intraday time series during those minutes (see the sudden and dramatic drop in the index visibile towards the end of the time series snapshot). What happened during those \textit{few} minutes?

\begin{figure}[!htbp]
\includegraphics[scale=0.7]{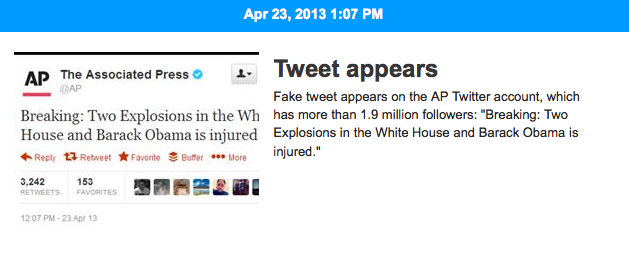}
\caption{AP tweet reporting explosions at the White House.}
\label{fig:faketweet}
\end{figure}

At 1:07 p.m.,  Associated Press tweeted that the US president had been reported injured in an explosion at the White House (see Fig. \ref{fig:faketweet}). The tweet was fake, probably due to a hacker that had temporary access to AP's account. The tweet was immediately retweeted a number of times (about 4000 in less than a minute, and even more in the next minutes). 

%

This two-minute storm, or ``flash-crash'' -- which led to a $\sim 1\%$ drop and a loss of \$130 billion in stock value -- cannot be explained by only \textit{one} tweet: indirect (network) effects have to be taken into account, probably also due -- as several observers have pointed out -- to the increasing adoption of high-frequency automated trading strategies. 

We believe that, in addition to this, AP's authority as an independent and reliable agency (coupled with the ``check mark'' that certifies a \textit{verified} Twitter account), led investors (indepedently from their strategy) to immediately \textit{trust the consequences} of such a tweet, without verifying its truthfulness. 

Whether or not the fake tweet was intentionally meant to destabilise the stock market is still an open question. Doubtless, it shows that the financial system is intrinsically vulnerable to \textit{opinion shifts}, regardless of the truthfulness of the underlying fact. The vulnerability of financial markets to such threats represents an open also for regulators as it leaves observers ``speculating once more about their vulnerability to breaking news in the age of social media''.\footnote{\textit{AP Twitter hack causes panic on Wall Street and sends Dow plunging}, The Guardian, April 23, 2013. Online version available at: \url{http://www.theguardian.com/business/2013/apr/23/ap-tweet-hack-wall-street-freefall}}

We have now come full circle. Referring, once more, to the Thomas theorem, \textit{the consequences were real}, although the actual ``situation'' was not.\footnote{Many scholars have pointed out that  \textit{self -- fulfilling prophecies}, i.e. statements or predictions that cause themselves to be true, represent a conundrum in the theory of social systems.  R. Merton's \textit{Social Theory and Social Structure} (1968) provides an extensive philosophical and historical portrayal of this fascinating concept.}

\section{Opinions and prices: a network model\label{sec:model}}

Consider a set $V$ of $i = 1, \ldots, n$ agents, one risky asset and one risk -- free asset (with interest rate $r$). The price dynamics is denoted by $(p(t))_{t \in \mathbb{N}}$. Let $z_i(t)$ be the quantity (number of shares) of the risky asset purchased by agent $i$ at time $t$. Finally, let $(y(t))_{t \in \mathbb{N}}$ be the dividend at $t$ (which will be assumed to be i.i.d. throughout this chapter). The wealth dynamical process for each agent can be described by the following equation (expressed in vector form):

\begin{equation}
\matr{w}(t+1) = (1 + r) \matr{w}(t) + \left(p(t+1) + y(t+1) - (1+r) p(t)\right) \matr{z}(t)
\end{equation}

In order to capture the opinion dynamics, we will denote the \textit{opinion of agent $i$ about the price and the dividend innovation} with the following expected value:

\begin{equation*}
x_i (t) = \mathbb{E}_i (p(t)) + \mathbb{E}_i (y(t)) = \mathbb{E}_i (p(t) + y(t))
\end{equation*}

The opinion at time $t$ is conditioned to the information set available at time $t-1$. Opinions are heterogenous for what regards the first moment of the distribution. The assumption can be extended to higher moments of the distribution but this case will not be explored here. The \textit{opinion about the variance} of wealth is constant and equal for all agents, in line with Hommes and Wagener (2005, review paper) and Chiarella \textit{et al.}. (2007).   The \textit{opinion} (or, in the originl context, \textit{belief}) that agent $i$ has about her wealth can be then described as follows:

\begin{eqnarray}
\mathbb{E}_i (w_i (t + 1)) &=& (1 + r) w_i(t) + x_i(t+1)z_i(t) - (1 + r) p(t) z_i(t) \nonumber \\ \smallskip
V_i (w_i(t + 1)) &=& \sigma^2 \;\; \;\;\; \forall i \in V, \forall t \in \mathbb{N} \nonumber 
\end{eqnarray}

\smallskip
\smallskip

Agents optimize their wealth in mean/variance, with the following utility function:

\begin{equation*}
u_i (w(t)) = -e^{-a_i W(t)},
\end{equation*}

where $a_i$ is the  Constant Absolute Risk Aversion (CARA) coefficient. This approach is consisent with the literature we refer to (Hommes \& Wagener, Chiarella \textit{et al}.). The wealth-maximisation problem at $t$ can be then expressed as follows:

\begin{eqnarray}
\max_{z_i(t)}&\left\{\mathbb{E}_i (w(t+1), t) - \frac{a}{2} \sigma^2\right\} \nonumber
\end{eqnarray}

thus leading to 

\begin{equation}
z_i(t) = \frac{1}{a\sigma^2} \left(x_i(t+1) - (1+r) p(t)\right)
\end{equation}

as the \textit{optimal} demand of the risky asset for agent $i$ at time $t$.
Hence, the optimal demand (number of shares) of agent $i$ at time $t$  proportional to the difference between the opinion on the risky asset price and dividend at$t+1$ the capitalized price at $t$, and inversely proportional to the absolute risk aversion coefficient $a$ and, obviously, the variance.

We assume that there is a limited \textit{outside supply per agent}, that will be denoted by $z^s$ (with $n z^s$ obviously being the overal outside supply). We have a convenient way to obtain the equilibrium price of the risky asset via the following equilibrium:

\begin{equation}
\frac{1}{n} \sum_{i = 1}^N \frac{1}{a \sigma^2} \left(x_i(t+1) - (1+r)p(t)\right) = z^s
\end{equation}

which finally leads to the following \textit{pricing equation}:
\begin{equation}
\label{eq:pricing}
(1+r) p(t) = \underbrace{\frac{1}{n} \sum_{i = 1}^N x_i(t+1)}_{\text{average opinion}} - \;  \underbrace{a\sigma^2 z^s}_{\text{risk premium}}
\end{equation}

As previously mentioned, we will assume that the dividend sequence $(y(t))_{t = 1, 2\ldots}$ is i.i.d., with variance $\sigma^2$ and will be the \textit{only source} of randomness in the model. If $\sigma = 0$, then price is deterministic and given only by the discounted average opinion.

\subsection{Introducing the network interaction}

We now  describe the opinion dynamics $\matr{x}(t)$ and the price dynamics $p(t)$ for the risky asset. Hommes and Wagener (2005) and, in more detail, Gaunersdorfer and Hommes (2007)  assume an interaction based on the success of the trading strategy basing it on a \textit{fitness function}. 

Our contribution in this work lies in adopting an \textit{explicit} interaction strategy, where agents learn from others according to specific rule. The interaction among agents, capture by a \textit{confidence network}, whose topology will be crucial in our model. Agents are modeled as vertices in the network. The network is time-dependent and evolves according to specific update rules that capture different situations in the financial system. The update rule modifies the underlying interaction network topology among agents.

Agents are \textit{boundedly rational} and DeGroot's approach is particularly suitable in our context because, ``incorporate indirect information in a boundedly rational way'' (cit. Golub and Jackson \cite{golub2010}).

Hence, we will assume a \textit{pairwise interaction}  with a dynamics of the kind described within the \textit{consensus problem} of DeGroot (1974). This issue was later tackled -for example- by Berger (1981) and, in a more general \textit{opinion dynamics} framework,  recently developed by Hegselmann and Krause (2005) and Lorenz (2007). The Methods Section of this paper provides with a detailed and rigorous discussion of these models and the mathematical background that will be needed in develop our model in rest of this Section. In the following paragraphs, we will briefly outline the main aspects and conclusions.

As mentioned in the Introduction, the basic problem of DeGroot was to develop a model to find a \textit{consensus} (i.e. a shared opinion) within a set of agents $V$  who revise their opinions by averaging (or ``'pooling') other agents' opinions. In particular, for each pair of agents $(i, j)$, agent $i$ assings a weight $a_{ij} \in [0, 1]$ to the opinion of agent $j$ ``to accommodate the information and expertise, the opinions and judgements, of the rest of the group''.

In this original statement of the problem, it is assumed that the linear combination is convex (i.e. $\sum_{j = 1} a_{ij} = 1$ and does not vary over time. As such, the model borrows some analytical results from Markov chain theory in order to find conditions for the convergence to a common opinion. A later, more general approach is the one of Hegselmann and Krause (2005), who introduce the  \textit{bounded confidence} opinion dynamics scheme: the weighing structure is time-dependent.

A key aspect in our model is the \textit{topological classification} of agents into \textit{essential} and \textit{inessential} classes. Intuitively, the set of agents can be split into subsets of strongly connected components (SCC). Agents within the same SCC \textit{communicate}, in that it is always possible to find a path from and to any pair of agents. In other words, within a SCC, agents have \textit{direct or indirect} positive confidence weight in any other agents within the same SCC. When no agent belonging to a specific SCC puts a positive confidence weight on an agent belonging to another SCC, then the first SCC corresponds to an \textit{essential} class. Vice versa, when one or more agents belonging to a specific SCC put a positive weight onto an agent belonging to another SCC, then the class is said to be \textit{inessential}. A rigourous formulation of this concept is provided in the Methods Section.

The relationship between essential and inessential classes, as pointed out by Berger (1981), will determine whether a common or fragmented opinion will be reached and how. Opinions of agents in the essential classes are considered by other inessential agents, but not vice versa. Predominant opinions and hence, prices, will then be determined by a linear combination of the opinions of the essential agents. 
Loosely speaking, essential agents are the \textit{opinion leaders} whereas inessential agents are \textit{opinion followers}. 

Our setup, however, allows that essential and inessential classes of agents may vary over time, according to the update rule. In other words, an agent can change her topological role in time. Since we are studying the model by means of simulations, we then need an algorithm to classify essential and inessential agents and find the class they belong to at each $t$. 

\smallskip

The \textit{opinion revision process} is captured by the following equation:

\begin{equation}
\label{eq:opinionrevisions}
\matr{x}(t) = \matr{A}(t) \; \matr{x}(t-1)
\end{equation}

which, coupled with Equation \ref{eq:pricing} will give the price at each $t$.

The process that we will assume in this model involves a change in the confidence network at each $t$, as a convex linear combination of the previous confidence matrix and an \textit{update matrix} $\matr{C}(t)$:

\begin{equation}
\label{eq:adaptive_strategy}
a_{ij}(t) = \alpha_i \; c_{ij}(t) + (1- \alpha_i) \; a_{ij}(t-1)
\end{equation}

\smallskip

where $\alpha \in [0, 1]$ is an \textit{update propensity} parameter. This parameter is key in our model because, as it will be shown in the simulations, the behaviour of the system depends non-trivially on it. The element $c_{ij}(t)$ of $\matr{C}(t)$ represents the update (revision) of the confidence weight that $i$ has towards $j$, and is determined according to the following rule:

\smallskip
\smallskip

\begin{equation*}
c_{ij}(t) := \left\{\begin{array}{cc}
\frac{1}{\# I(i, \matr{x}(t-1), p(t-1))} & \mbox{if } j \in I(i, \matr{x}(t - 1), p(t-1)) \\
0 & \mbox{otherwise}
\end{array} \right.
\end{equation*}

\smallskip

It is immediate to prove that, according to this rule, $\matr{C}(t)$ is row stochastic $\forall t$ and, therefore, also $\matr{A}(t)$ is also row -- stochastic $\forall t$.

The model is non - linear because of the interaction among the opinions, prices and the confidence network. In the context of consensus/opinion dynamics, Hegselmann and Krause (2005) also  refer to this kind of model as \textit{non--linear} and point out that ``the most difficult type of model occurs if the weights depend on opinions itself because then the model turns from a linear one to a \textit{non--linear} one''.

Differently from the standard approach, we do not focus on specific \textit{trader types}. Rather, agents are classified according to the role they play in the topological structure, hence allowing to capture their beahviour along a  \textit{continuum} of possible strategies, coupled with the evolving topology of interactions. The choice of the set $I(i, \matr{x}(t-1), p(t-1))$ for each agent $i$ will lead to different types of models and will be now detailed.


\subsubsection{Bounded confidence model}

In the first case, we will adopt a similar approach to the one in Hegselmann and Krause (2005) and reported in Equation \ref{eq:boundedconfidence} in the Methods section. The authors refer to it as \textit{bounded confidence} in that a threshold exists when determining to whom an agent puts a confidence weight. We also label this model as \textit{bounded confidence} (BC). The adaptive update matrix $\matr{C}(t)$ is:

\begin{equation}
\label{eq:bounded}
I(i, \matr{x}(t-1), p(t-1)) = \{j \mbox{ s.t. } |x_i (t-1) - x_j (t-1) | \leq \epsilon_i\}
\end{equation}

\smallskip

In this scenario, there is no need to look at \textit{fundamentals} in order to understand the price dynamics. Prices will be driven only by the opinions and the opinions will be pooled according the bounded confidence non -- linear model. The non  -- linearity in the model will lead to opinion polarization and fragmentation, thus the average in Eq. \ref{eq:pricing} can increase or decrease dramatically according to such level of polarisation. It is important to notice that $\{i\}$ is included in the set $I(i, \matr{x}(t-1), p(t-1))$ by construction.

\subsubsection{Price adaptive strategy}

The idea in this setting is that those agents whose opinions are the closest to the actual price at time $t$ (price adaptive, PA) will be followed, in the next time step, by other agents. In an ``adaptive'' perspective, these agents are thus those capable of understanding the nontrivial impact of \textit{both} opinions on price and the impact of dividend innovations.

In this adaptive setting, agents revise the confidence matrix as in Equation \ref{eq:adaptive_strategy}, but the set $I(i, \matr{x}(t-1), p(t-1))$ is now given by:

\begin{equation}
\label{eq:priceadaptive}
I(i, \matr{x}(t-1), p(t-1)) = \{i\} \cup \left\{j \mbox{ s.t. } \left| \frac{p(t-1) -  x_i (t-1)}{p(t-1)}\right| \leq \epsilon_i\right\}
\end{equation} 

\smallskip

In other words, in this case, $c_{ij}(t)$ represents an adaptive strategy where agents tend to follow more agents that had close opinions to the actual realized price. If $\epsilon_i$ is a nonzero constant $\forall i \in V$, then $\matr{C}(t)$ is a rank one matrix at each $t$. Since the price is driven both by the opinion dynamics and an exogenous divided innovation, this model reflects a situation where agents want to follow other agents that have proven to be successful in the past.

The reason why we insert the set $\{i\}$ in the set $I$ in the set \ref{eq:priceadaptive} is because we want to have a non-empty set and also allow the matrix $\matr{A}(t)$ to have a positive diagonal (also in the case $\alpha = 1$).

\subsubsection{Fundamental benchmark}

The reversion of asset prices towards a fundamental value is a long--debated issue in the financial literature.\footnote{References include Lo and Mackinlay (1988), who develop a specification test and claim that ``prices do not follow random walks''; Fama and French (1988); and Poterba and
Summers (1988) who focus explicity on mean reversion in asset pricess, tackled by Stefani \textit{et al.} (2010) in commodity trading.}
When all agents have identical long -- term expectations (opinions) on the price, i.e. $x_i(t) = c, \forall i \in V$, we can assume  $\matr{x}(t)$ to be the analogous of a consensus limit vector where all agents belong to the same essential class. The \textit{pricing equation} \ref{eq:pricing} can be then rewritten as:

\begin{equation*}
(1 + r) p(t) = c - a\sigma^2z^s
\end{equation*}

We can also define a \textit{fundamental price} given by the discounted sum of all expected future dividends:

\begin{equation}
\label{eq:fundamental}
p^* (t) = \sum_{k = 1}^{\infty} \frac{\mathbb{E}(y_{t + k}, t) - a\sigma^2 z^2}{(1 + r)^k}
\end{equation}

which can be set in an analogous way of a perpetuity with interest $r$. An interesting aspect is that, at a specific time $t$, should the confidence matrices be fixed, one can assume that there are $g$ different asymptotic opinions, each of which can be associated to an \textit{essential class} of the kind described in the Methods section.  When $(y(t))_{t = 1, 2, \ldots}$, then $\mathbb{E}[y_t] = \bar{y}$ and the fundamental price is constant over $t$:

\smallskip

\begin{equation*}
p^* = \sum_{k = 1}^\infty \frac{\bar{y} - a\sigma^2 z^s}{(1+r)^k} = \frac{\bar{y} - a \sigma^2 z^s}{r}
\end{equation*}

\smallskip

We will show, by simulations, that in this model based on fundamentals (FB), opinion shifts in the essential classes will lead to significant changes in prices.

In this case, we assume the set $I$ to be determined as follows:
\smallskip
\smallskip

\begin{equation*}
I(i, \matr{x}(t-1), p(t-1)) = \{i\} \cup \left\{j \mbox{ s.t. } \left|\frac{p^* - x_i (t-1)}{p^*}\right| \leq \epsilon_i\right\}
\end{equation*}

\smallskip
\smallskip

This implies that all agents will see deviations from the fundamental as temporary and will tend to trust agents whose opinion does not deviate significantly from the fundamental. In behavioral terms, agents want to ``find a confirmation'' from other agents that the fundamental is the correct long-term benchmark. However, small and temporary deviations are perceived as reasonable and agents will still try to maximize their short - term wealth (myopic maximisation) by exploiting them.

There is one general case when the fundamental price is achieved in ``aggregate'' at time $t$, i.e. when the \textit{average opinion} on the price is equal to the fundamental ($\frac{1}{n}\sum_{i =1}^n x_i(t) = p^*(t)$). As a particular case, obviously, when at $t$, all the agents' opinions are all equal to the fundamental at time $t$.

The first, more general, case could occur in a number of situations. For example, even in presence of high variability in the opinions, the system can reach the fundamental price anyway, if the average opinion coincides with it.

%

\subsection{Combinations of strategies}

The flexibility of the model we propose allows to have different strategies (BC, PA, FB) for each agent; different trader types  (e.g. fundamentalists, chartists, contrarians, etc.) can be modeled within this framework.
Among the many possible combinations, an interesting analysis would be to assume different strategies for essential and inessential agents assuming, for instance, two interesting scenarios that will be the subject of future research and will not be explored here. 

In the first scenario, essential (opinion drivers) agents' behaviour would be encapsulated within a Bounded Confidence model with a very low update propensity, whereas inessential agents (opinion followers) would be following a Price Adaptive strategy with a very high update propensity. This would lead to a situation where a few leaders drive the opinion dynamics and many followers adjust very quickly to those agents whose opinion was the most accurate at each time step.

The second scenario is analogous to the first, with a significant difference. The essential agents are split into two categories: opinion leaders who follow a fundamentalist strategy with a very low update propensity and other ``impulsive'' essential agents who follow a price adaptive strategy. In this case, the essential agents in the second category would enhance little fluctuations when price temporarily depart from the fundamentals.

Although not explicitly explored here, the model potentially allows to capture agents who explicitly want to be followed (thus determining the market), acting as opinion leaders. This obviously introduces the problem of the controllability of the confidence network. The framework might then be useful also for regulators.

\section{Simulation results and discussion}

In this section, we show and discuss simulation results on the dynamics described by Equations \ref{eq:pricing}, \ref{eq:adaptive_strategy} according to the three models (BC, Equation \ref{eq:bounded}; PA \ref{eq:priceadaptive} and FB, \ref{eq:fundamental}). We start with the analysis of the price dynamics and later provide some descriptive statistics on the distribution of returns.
\noindent
Given the number of parameters in the model, we will make the following simplifying  assumptions:

\begin{enumerate}
\item The variance on the dividend innovations is constant  $\forall t$ and  $\forall i$ (i.e. on the innovation opinions) and equal to $\sigma$;
\item $\mathbb{E}_{it}(t_{t+1}) = \bar{y}$, $\forall i, t$;
\item in any case, $\epsilon_i = \epsilon$, $\forall i$.
\end{enumerate}

The first part of the simulation analysis (Figures \ref{fig:BC_sims}, \ref{fig:FB_sims} and \ref{fig:PA_sims}) reports a representation of the price $p(t)$ in time, obtained from the pricing equation \ref{eq:pricing}, by averaging out  $1000$ realizations of the process in Equation \ref{eq:adaptive_strategy} from the \textit{same} starting opinion profile $\matr{x}(0)$, that it is distributed as a log -- normal with mean $3$. The number of agents is set to $n = 100$.

In particular, figures \ref{fig:BC_sims}, \ref{fig:FB_sims}, \ref{fig:PA_sims} show simulations for the price dynamics for the models BC, FB and PA respectively. Some descriptive statistics are shown in Tables \ref{T:BCsk} -- \ref{T:FBku} for the price returns $r(t) = p(t)/p(t-1) - 1$. These statistics and the price dynamics reflect some well--known stylized facts in financial markets (e.g., clustered volatility, fat tails, asymmetric returns). In particular, although the return distributions are slightly asymmetric, we used the excess kurtosis as a way to compare tail distribution with those of a normal distribution one can see that fat tails arise in the BC non -- linear model (where polarisation occurs) when the update propensity parameter $\alpha$  is very high. This means that the quicker the fragmentation, the quicker non -- linear phenomena arise and fat tails occur. The PA model, on the contrary, produces no fat -- tails, as the result of a strong reversion towards the fundamental.

\newcommand{\scvalue}{0.6}
\begin{figure}[!htbp]
\centering
                \includegraphics[scale=\scvalue]{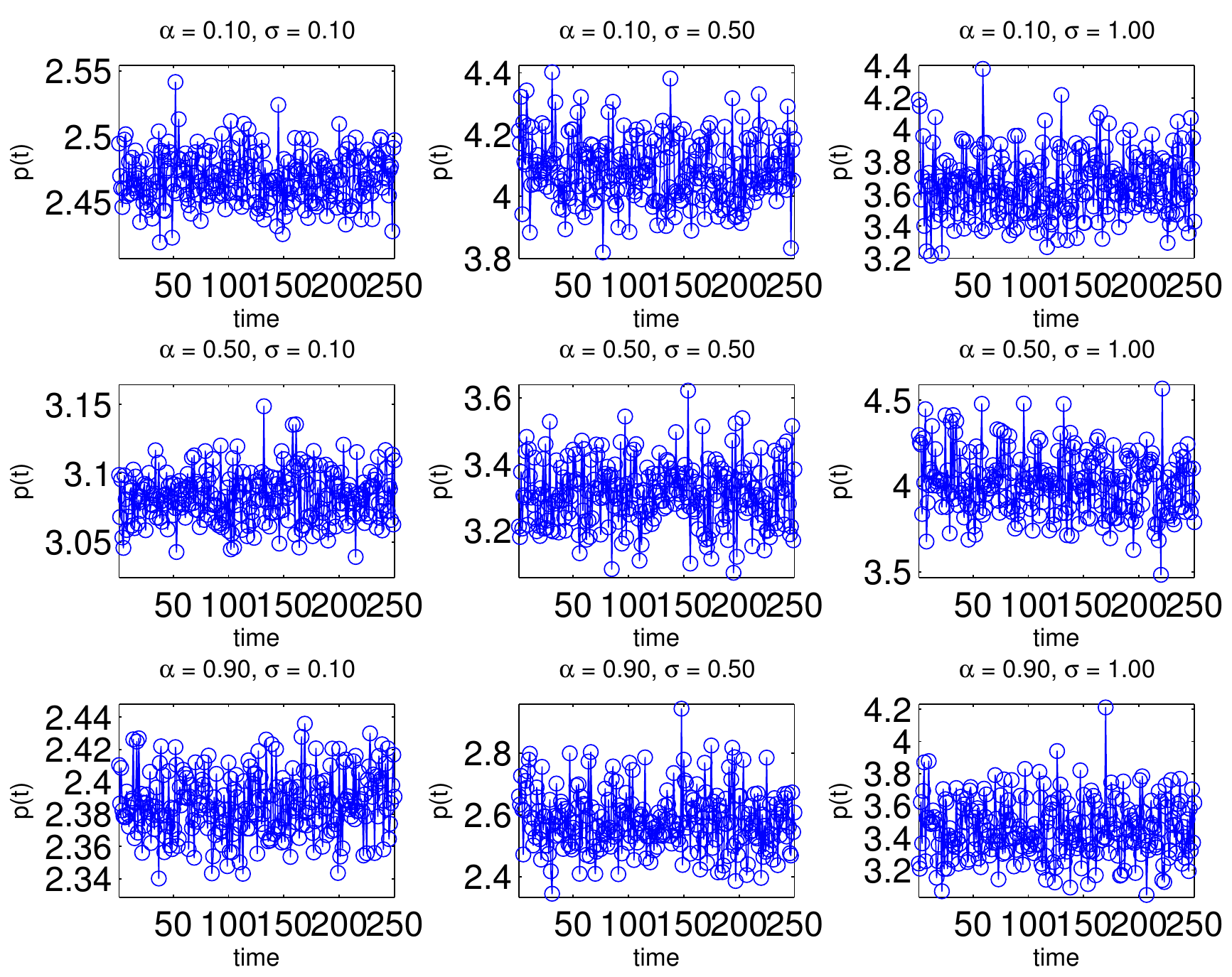}
                \caption{BC model for different values of $\alpha$ and $\sigma$.}
                \label{fig:BC_sims}
                \end{figure}

\begin{figure}[!htbp]
\centering
                \includegraphics[scale=\scvalue]{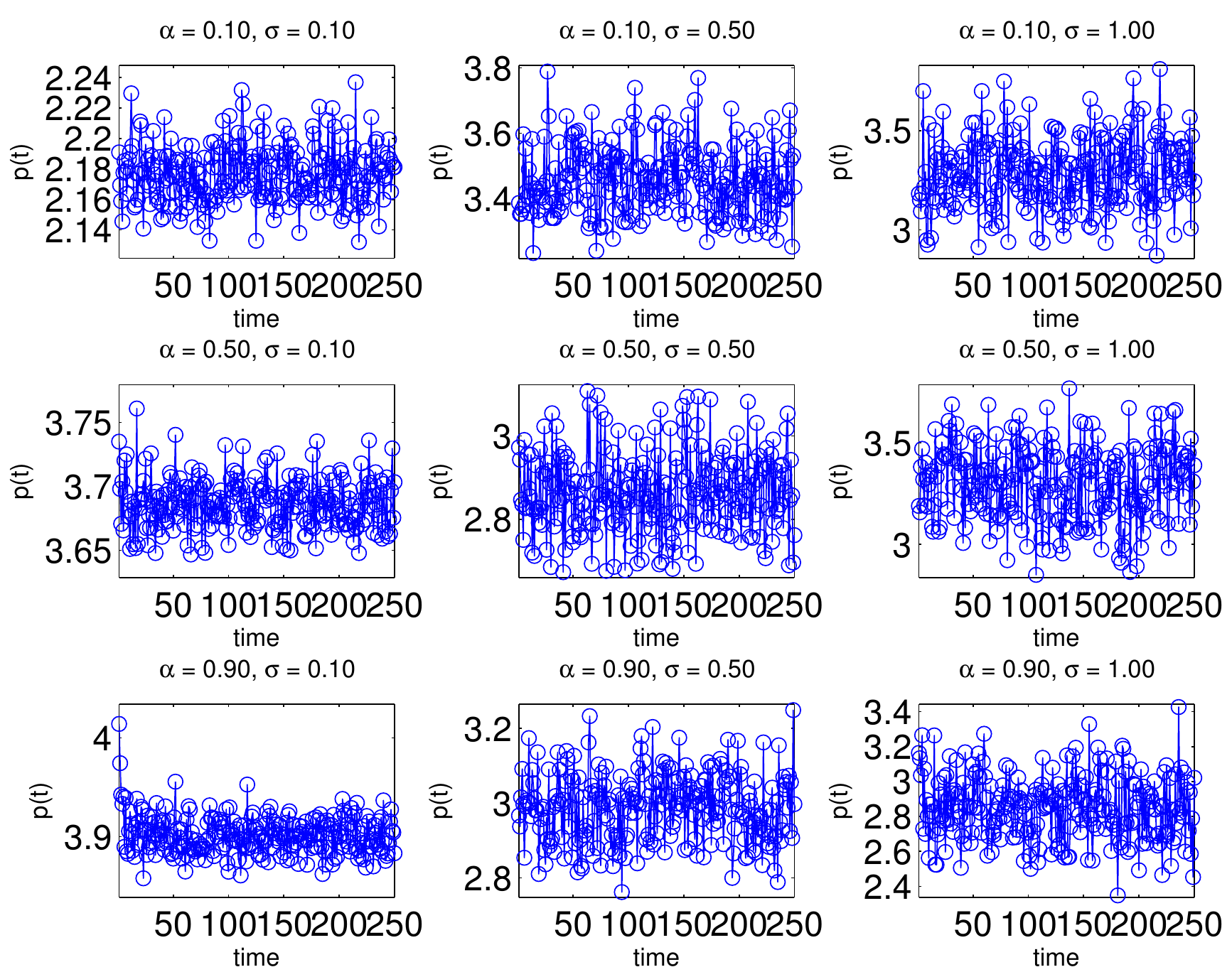}
                \caption{PA model for different values of $\alpha$ and $\sigma$.}
                \label{fig:PA_sims}
                \end{figure}

\begin{figure}[!htbp]
\centering
                \includegraphics[scale=\scvalue]{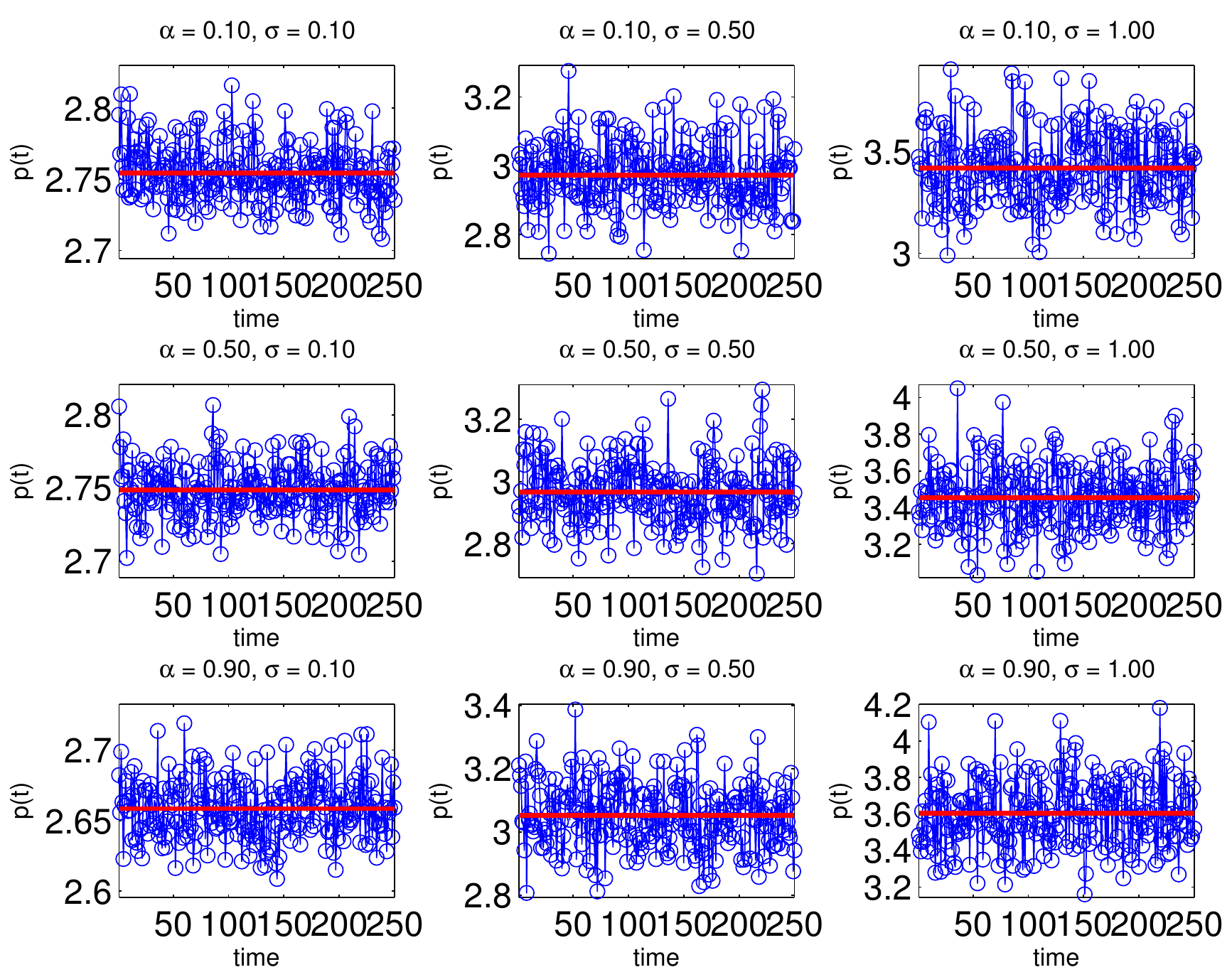}
                \caption{FB model for different values of $\alpha$ and $\sigma$. Fundamental value is the horizontal red line.}
                \label{fig:FB_sims}
                \end{figure}

\pagebreak

\begin{table}
\begin{center}
\caption{Descriptive statistics for the price returns $p(t)$ over $1000$ simulations for the three models and different values of the update propensity parameter $\alpha$ and $\sigma$.}\end{center}
\begin{subtable}[tbp]{0.5\textwidth}
\centering
\begin{footnotesize}\begin{tabular}{|l|c|c|c|}
\hline
&\textbf{$\sigma = 0.1$}&\textbf{$\sigma = 0.5$}&\textbf{$\sigma = 1$}\\\hline
\textbf{$\alpha = 0.1$}&0.20&-0.03&0.02\\\hline
\textbf{$\alpha = 0.5$}&-0.21&-0.16&0.25\\\hline
\textbf{$\alpha = 0.9$}&0.12&0.05&-0.02\\\hline
\end{tabular}
\end{footnotesize}
\caption{BC model -- skewness}
\label{T:BCsk}
\end{subtable}
\bigskip
\begin{subtable}[tbp]{0.5\textwidth}
\centering
\begin{footnotesize}\begin{tabular}{|l|c|c|c|}
\hline
&\textbf{$\sigma = 0.1$}&\textbf{$\sigma = 0.5$}&\textbf{$\sigma = 1$}\\\hline
\textbf{$\alpha = 0.1$}&-0.38&-0.45&-0.18\\\hline
\textbf{$\alpha = 0.5$}&-0.00&0.42&0.00\\\hline
\textbf{$\alpha = 0.9$}&0.29&-0.08&0.59\\\hline
\end{tabular}
\end{footnotesize}
\caption{BC -- (excess) kurtosis}
\label{T:BCku}
\end{subtable}

\begin{subtable}[tbp]{0.5\textwidth}
\centering
\begin{footnotesize}\begin{tabular}{|l|c|c|c|}
\hline
&\textbf{$\sigma = 0.1$}&\textbf{$\sigma = 0.5$}&\textbf{$\sigma = 1$}\\\hline
\textbf{$\alpha = 0.1$}&0.14&-0.02&-0.09\\\hline
\textbf{$\alpha = 0.5$}&0.09&-0.16&-0.13\\\hline
\textbf{$\alpha = 0.9$}&0.08&0.11&0.12\\\hline
\end{tabular}
\end{footnotesize}
\caption{PA model - skewness}
\label{T:PAsk}
\end{subtable}
\bigskip
\begin{subtable}[tbp]{0.5\textwidth}
\centering
\begin{footnotesize}\begin{tabular}{|l|c|c|c|}
\hline
&\textbf{$\sigma = 0.1$}&\textbf{$\sigma = 0.5$}&\textbf{$\sigma = 1$}\\\hline
\textbf{$\alpha = 0.1$}&0.10&-0.34&-0.14\\\hline
\textbf{$\alpha = 0.5$}&0.66&-0.22&-0.01\\\hline
\textbf{$\alpha = 0.9$}&-0.14&-0.12&-0.26\\\hline
\end{tabular}
\end{footnotesize}
\caption{PA model -- (excess) kurtosis}
\label{T:PAku}
\end{subtable}

\begin{subtable}[tbp]{0.5\textwidth}
\centering
\begin{footnotesize}\begin{tabular}{|l|c|c|c|}
\hline
&\textbf{$\sigma = 0.1$}&\textbf{$\sigma = 0.5$}&\textbf{$\sigma = 1$}\\\hline
\textbf{$\alpha = 0.1$}&0.05&-0.18&0.12\\\hline
\textbf{$\alpha = 0.5$}&-0.02&0.19&-0.17\\\hline
\textbf{$\alpha = 0.9$}&0.18&-0.00&0.08\\\hline
\end{tabular}
\end{footnotesize}

\caption{FB model - skewness}
\label{T:FBsk}
\end{subtable}
\bigskip
\begin{subtable}[tbp]{0.5\textwidth}
\centering
\begin{footnotesize}\begin{tabular}{|l|c|c|c|}
\hline
&\textbf{$\sigma = 0.1$}&\textbf{$\sigma = 0.5$}&\textbf{$\sigma = 1$}\\\hline
\textbf{$\alpha = 0.1$}&-0.09&0.23&-0.04\\\hline
\textbf{$\alpha = 0.5$}&-0.21&0.40&0.20\\\hline
\textbf{$\alpha = 0.9$}&-0.22&0.59&-0.39\\\hline
\end{tabular}
\end{footnotesize}
\caption{FB model - (excess) kurtosis}
\label{T:FBku}
\end{subtable}
\end{table}

\pagebreak

\clearpage

\subsection{Opinion shifts}
This specific simulation setup aims at explaining sudden drops in price followed by quick recoveries in presence of herding behaviour. In particular, we will try to assess whether the drop in the DJ30 following the fake tweet by AP can be explained in terms of network effects. To do so, we assume a ``sudden'' exogenous change at $t = 50$ for either some essential classes or the union set of all inessential classes. The basic idea is to assess whether changes in the opinion profiles of essential classes might have a different impact with respect to change in the opinion profiles of inessential classes, \textit{regardless of convergence}. In fact, the conditions outlined in the Methods Section show that essential classes play a role in determining the value towards which opinions converge, but very little can be said, from an analytical perspective, about short--term behaviour. In particular, we will see that changes in inessential classes do have an impact on slowing the recovery.

\begin{figure}[htbp]
\centering
\includegraphics[scale=1, trim = 0.0 0.0 3cm 0cm]{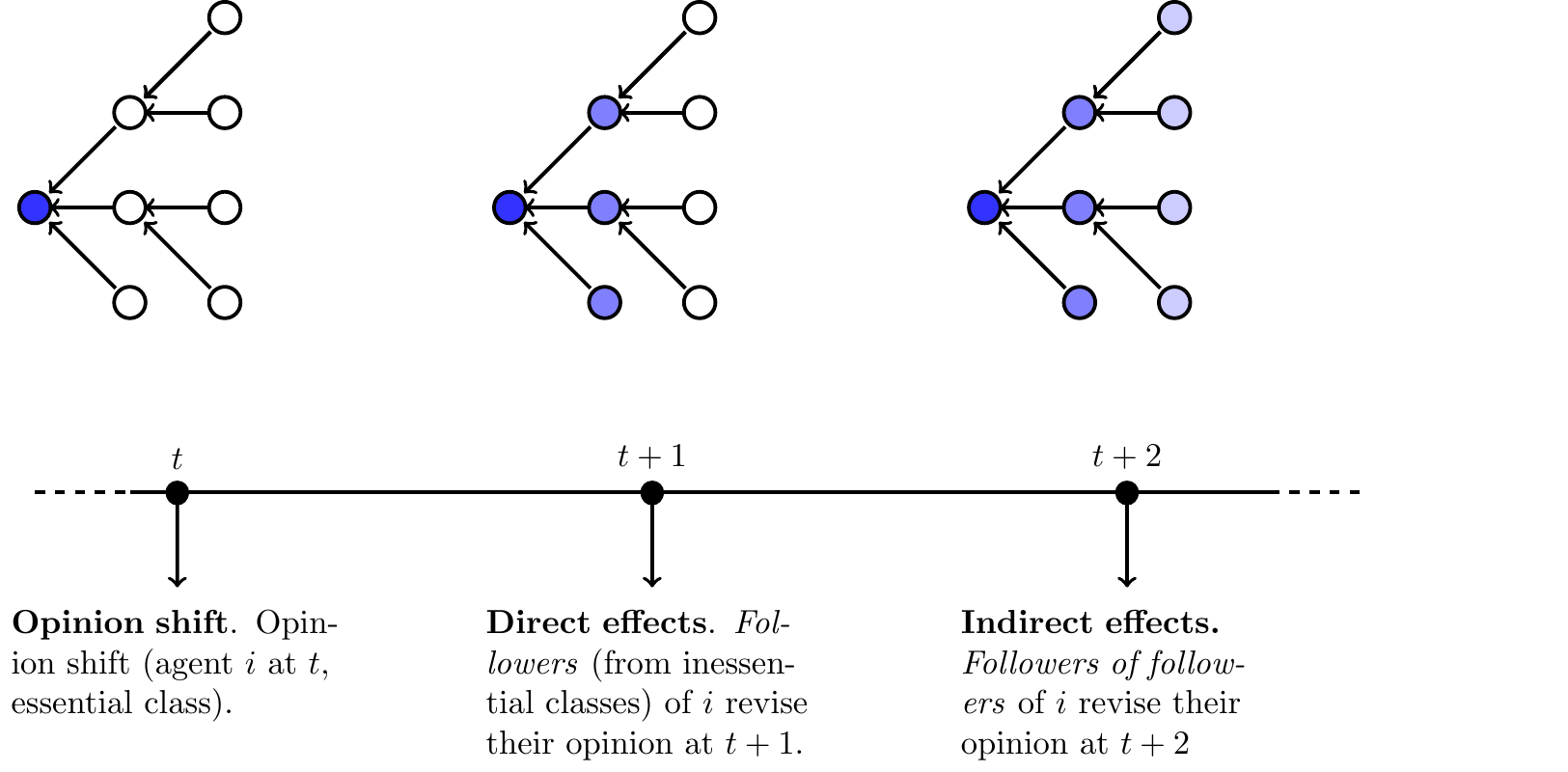}
\caption{Illustration of the opinion shift in an essential class and its reverberations on the inessential classes. Agent $i$ represents an essential class (the only one present in the graph) on her own.}\label{fig:explshift}
\end{figure}

Figure \ref{fig:explshift} provides a graphical illustration of the dynamics we assume in this simulation setup. The Figure reports a topology with one essential class (the node on the left hand side), and three inessential classes (in the center and on the right hand side). An opinion shift occurs at a specific time $t$ in an essential class composed of only one agent, producing a limited impact on the price. At $t+1$, one  agents for each inessential class  starts witnessing the shift and updates her opinion, the impact on the price therefore starts to increase, now involving four agents. At $t+2$, agents in two inessential classes update their opinions. The shift has now had an impact on eight agents, with a more significant impact on the price. 
n price but it will persist less. Should a \textit{revision} of the previous shift occur in the essential agent, this would take two time steps to lead the price to a full recovery.

\begin{figure}[!htbp]
\centering
                \includegraphics[scale=\scvalue]{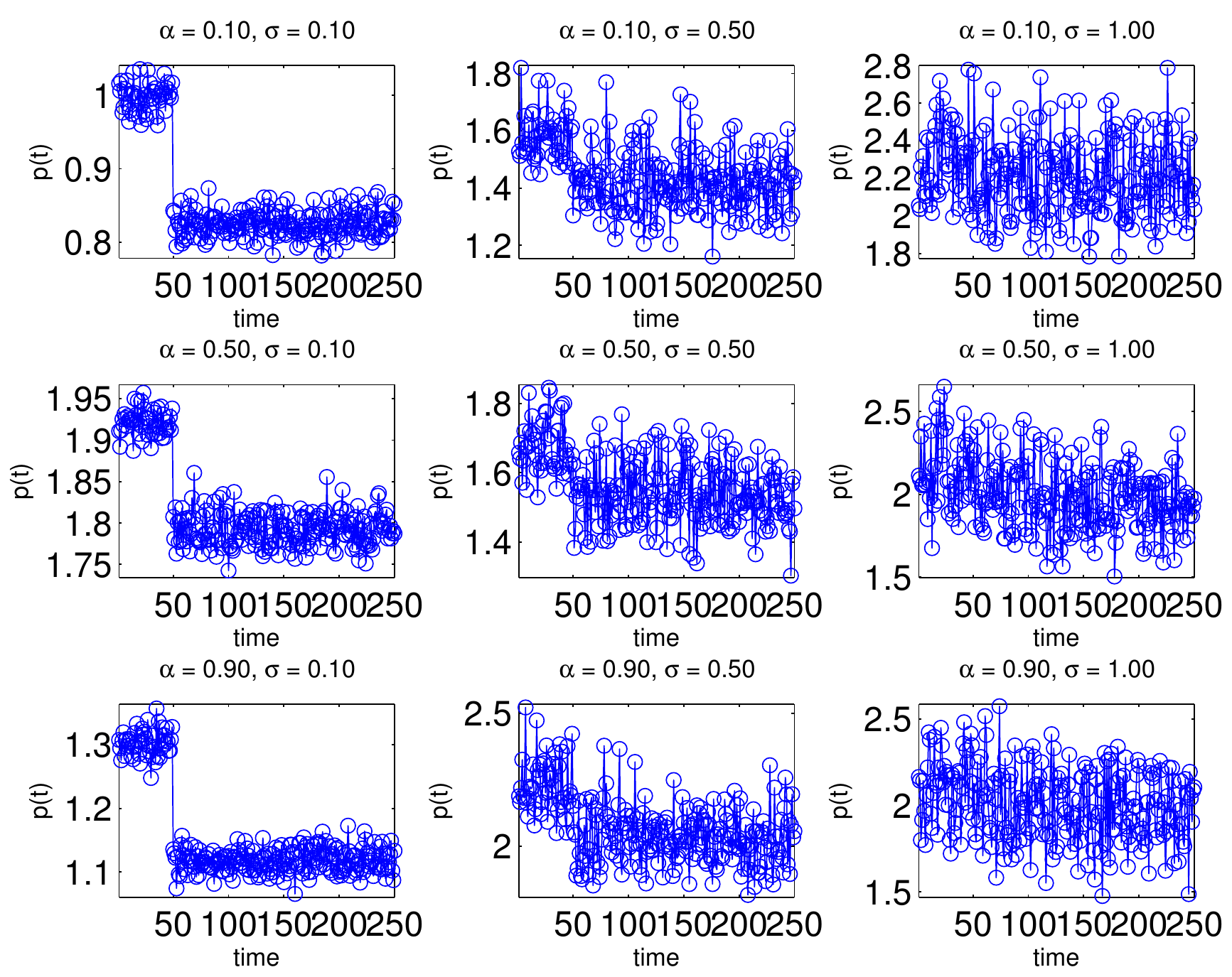}
                \caption{BC, opinion shift (drop) in essential classes}
                \label{fig:dropBCess}
                \end{figure}

\begin{figure}[!htbp]
\centering
                \includegraphics[scale=\scvalue]{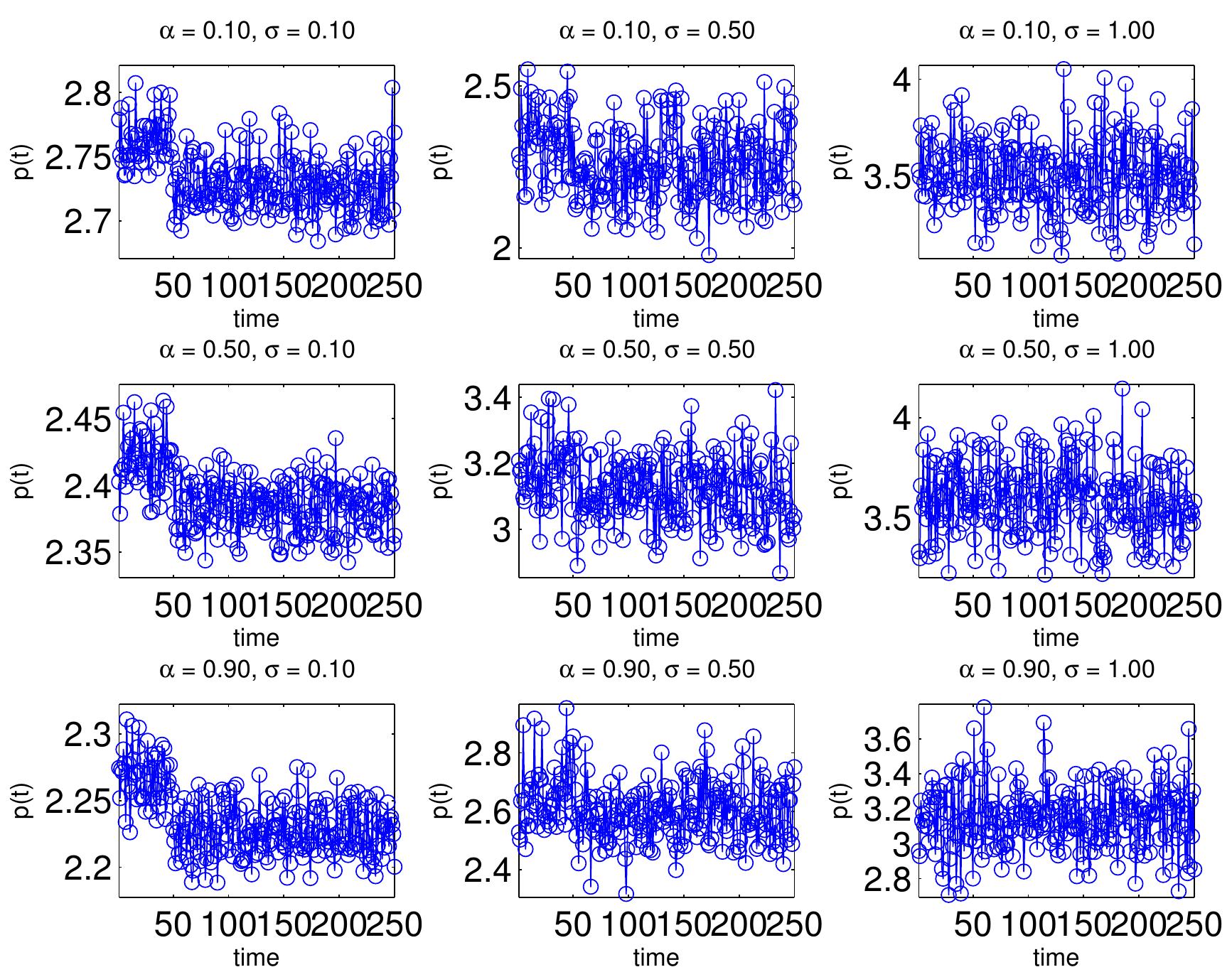}
                \caption{ BC, opinion shift (drop) drop in inessential classes}
                \label{fig:dropBCiness}
                \end{figure}

\begin{figure}[!htbp]
\centering
                \includegraphics[scale=\scvalue]{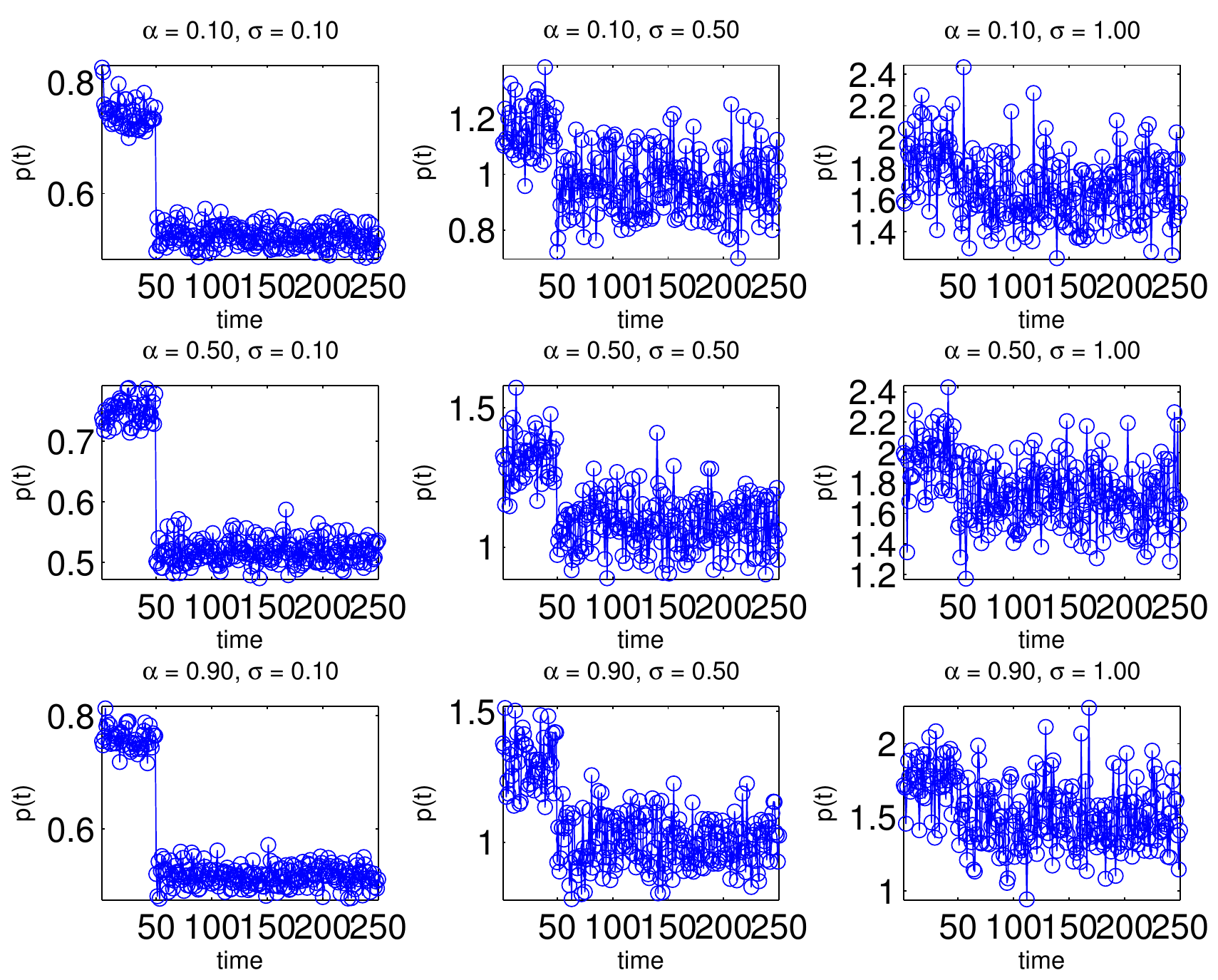}
                \caption{PA, opinion shift (drop) in essential classes}
                \label{fig:dropPAess}
                \end{figure}

\begin{figure}[!htbp]
\centering
                \includegraphics[scale=\scvalue]{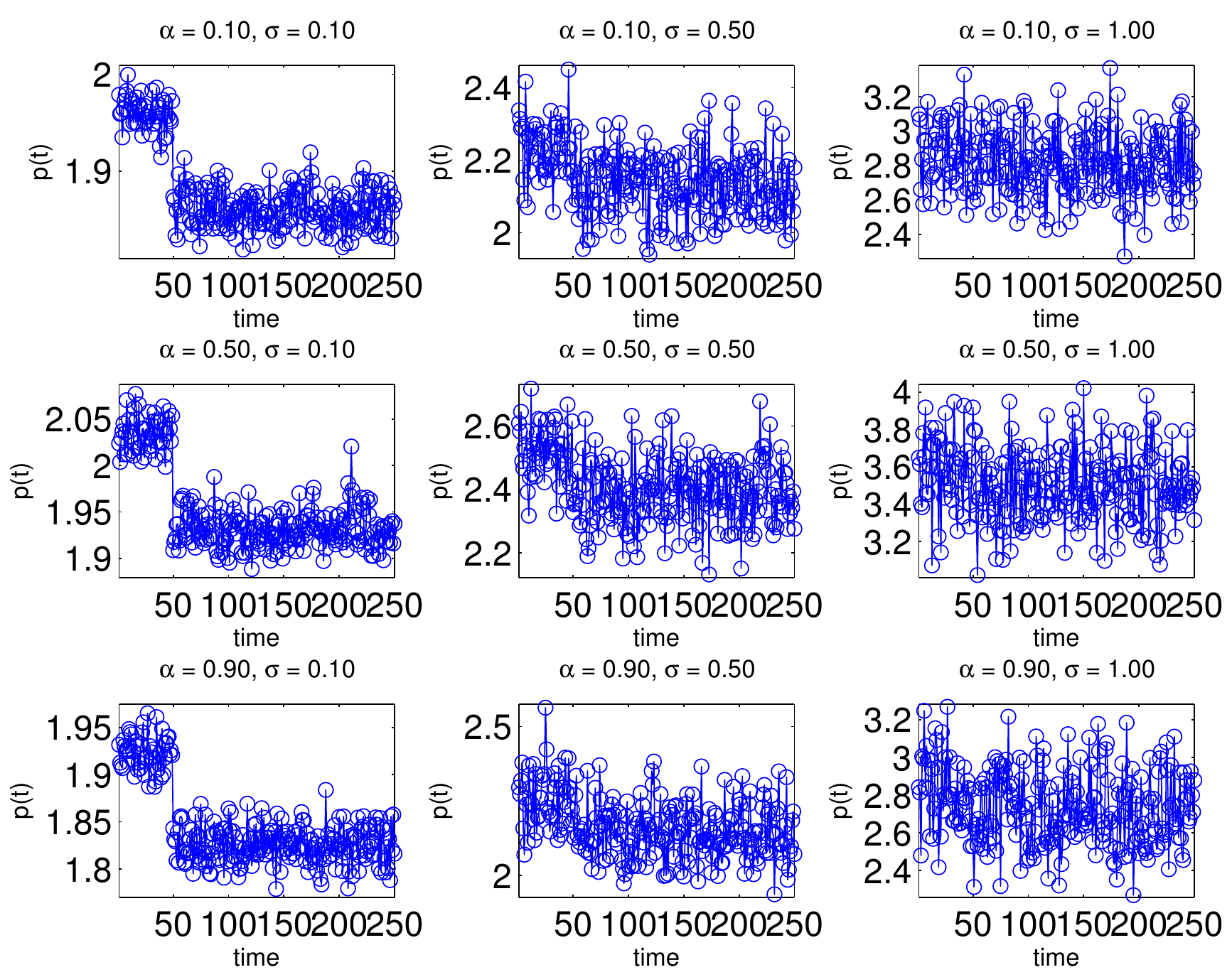}
                \caption{PA, opinion shift (drop) in inessential classes}
                \label{fig:dropPAiness}
                \end{figure}

Figure \ref{fig:dropBCess} shows that, in the BC model, when drops occur in the essential classes, they tend to persist for low values of $\sigma$: in fact, the \textit{bounded} nature of the model will provide a polarisation versus a low market scenario rather than a recovery. This is not the case when the drop occurs in the inessential classes (Fig. \ref{fig:dropBCiness}), where we observe a milder impact. Figures \ref{fig:dropPAess} and \ref{fig:dropPAiness} shows analogous findings for the PA model. The most interesting case is the FB model, where it is important to understand whether a return to the fundamental benchmark will occur and the time to recovery. 

\begin{figure}[!htbp]
\centering
                \includegraphics[scale=\scvalue]{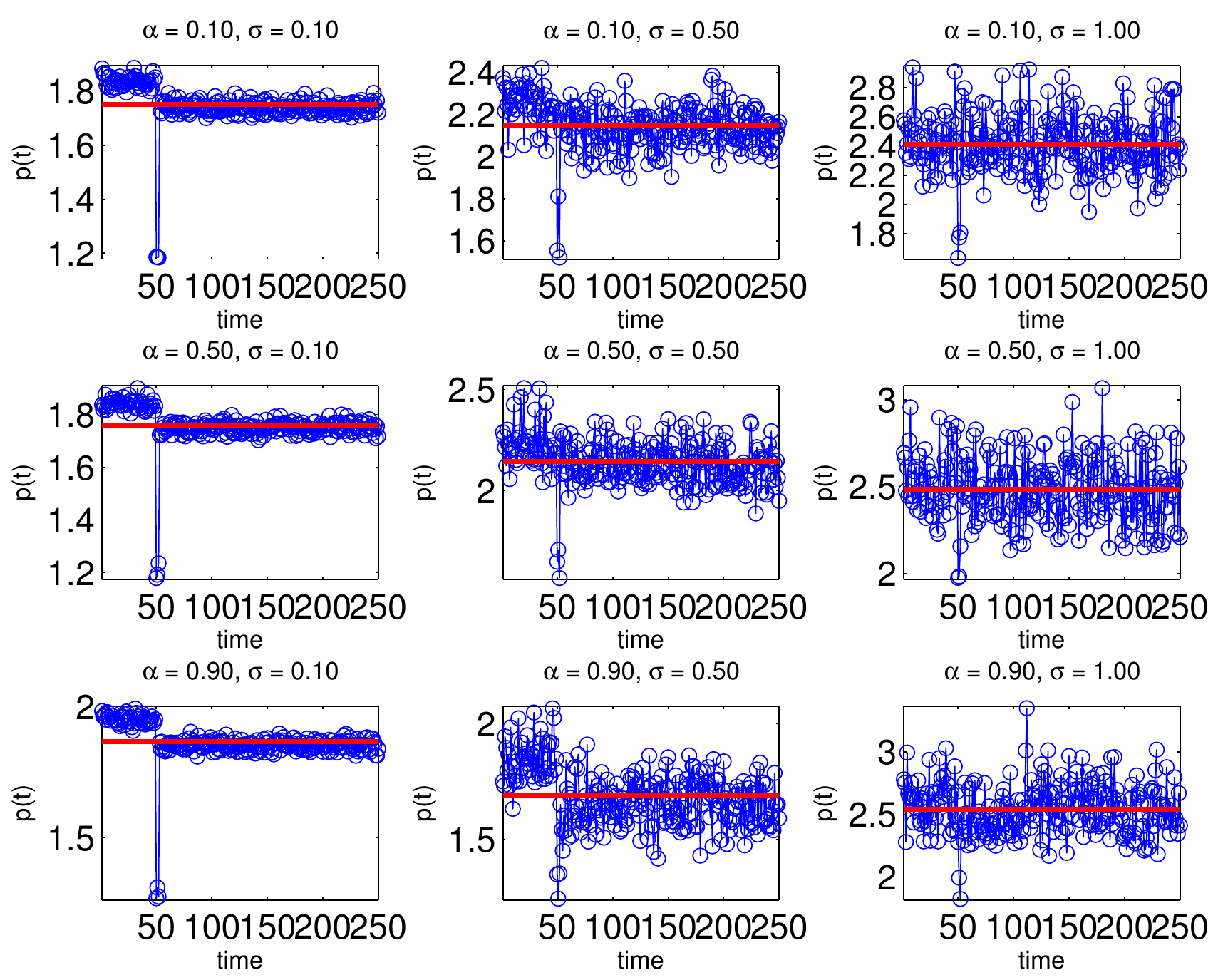}
                \caption{FB, opinion shift (drop) in essential classes. The fundamental value is in red.}
                \label{fig:dropFBess}
                \end{figure}

\begin{figure}[!htbp]
\centering
                \includegraphics[scale=\scvalue]{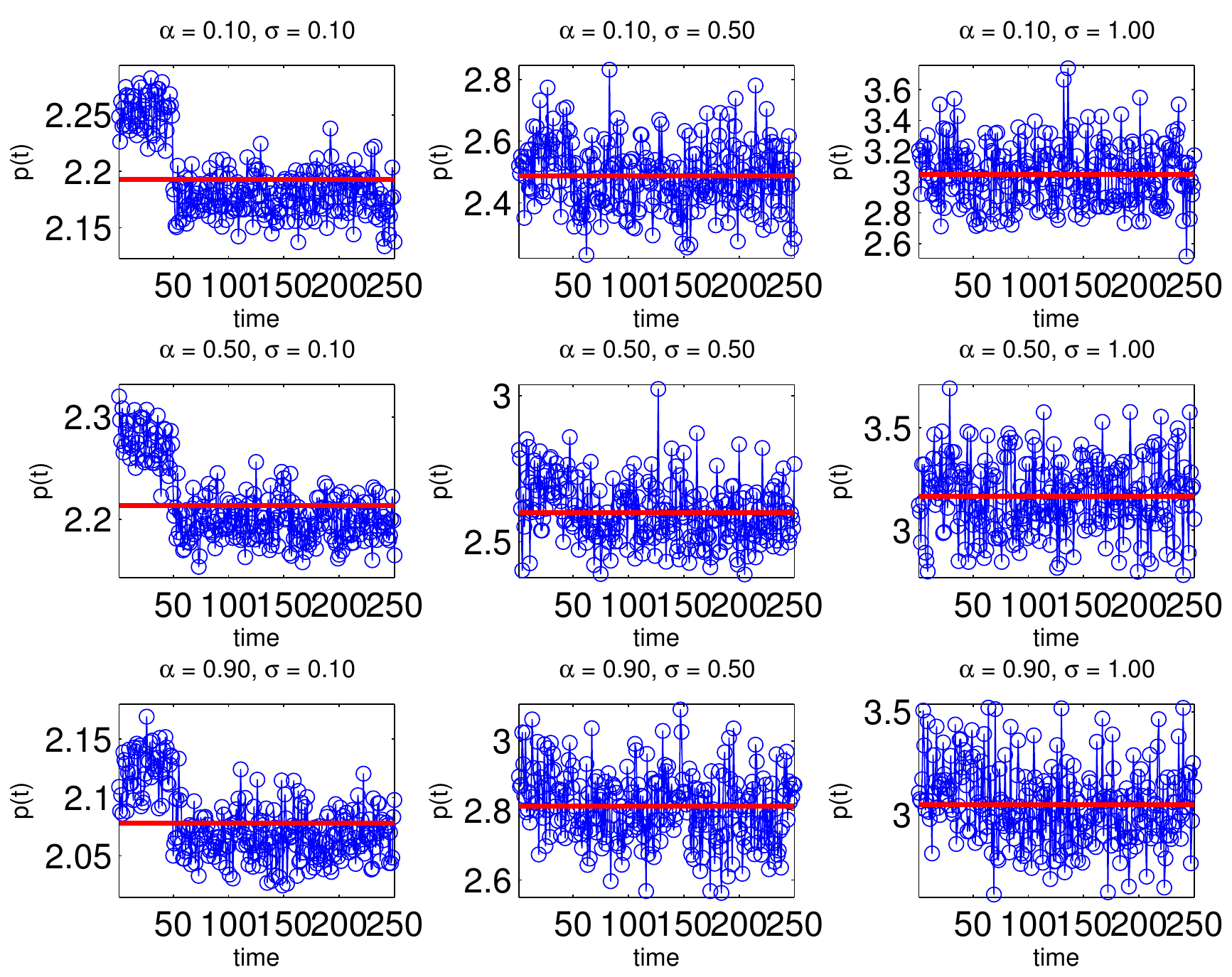}
                \caption{FB, opinion shift (drop) in inessential classes. The fundamental value is in red.}
                \label{fig:dropFBiness}
                \end{figure}

Figure \ref{fig:dropFBess} and  \ref{fig:dropFBiness} outline an important fact: drops happen and persist when opinion shifts occur in both essential and inessential classes, then recovering quite rapidly. However, the drops are much more pronounced in the essential class case and for \textit{lower} values of $\sigma$, thus showing that a higher variability in dividend innovations might lead to a more stable price dynamics w.r.t. sudden opinion shifts. The role of the update propensity parameter $\alpha$ is important for higher values of $\sigma$ in that lower propensity implies a lower degree of abandoment of previously formed opinions on the deviations from the fundamental, hence making recovery faster.

\bigskip

\newpage

\begin{figure}
        \centering
        \begin{subfigure}[b]{0.4\textwidth}
                \includegraphics[width=\textwidth]{dj_crash.jpg}
                \caption{ }
                \label{fig:many_dropsa}
        \end{subfigure}%
        \\
        \begin{subfigure}[b]{0.4\textwidth}
                \includegraphics[width=\textwidth]{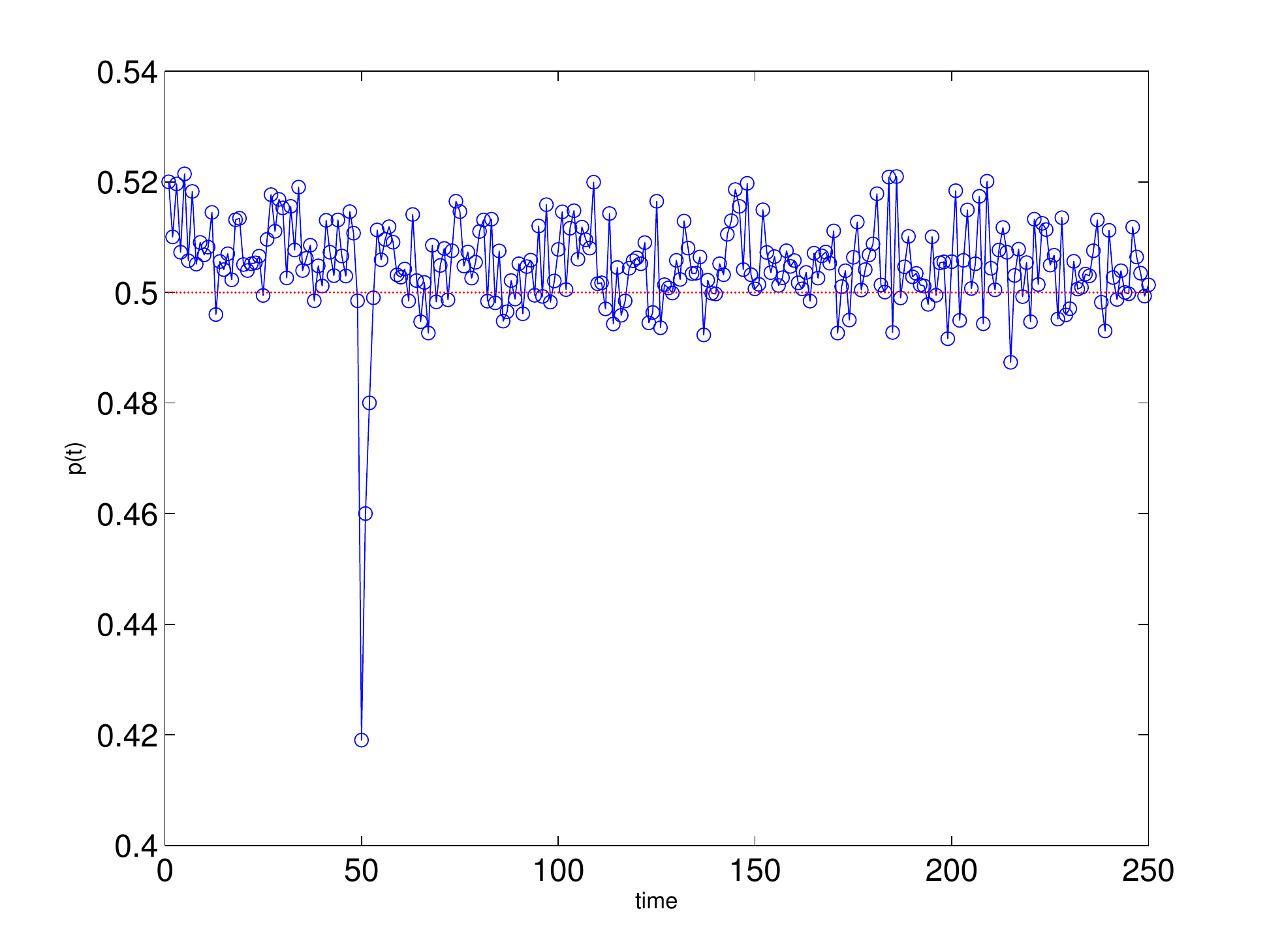}
                \caption{ }
                \label{fig:many_dropsb}
        \end{subfigure}%
        \begin{subfigure}[b]{0.4\textwidth}
                \includegraphics[width=\textwidth]{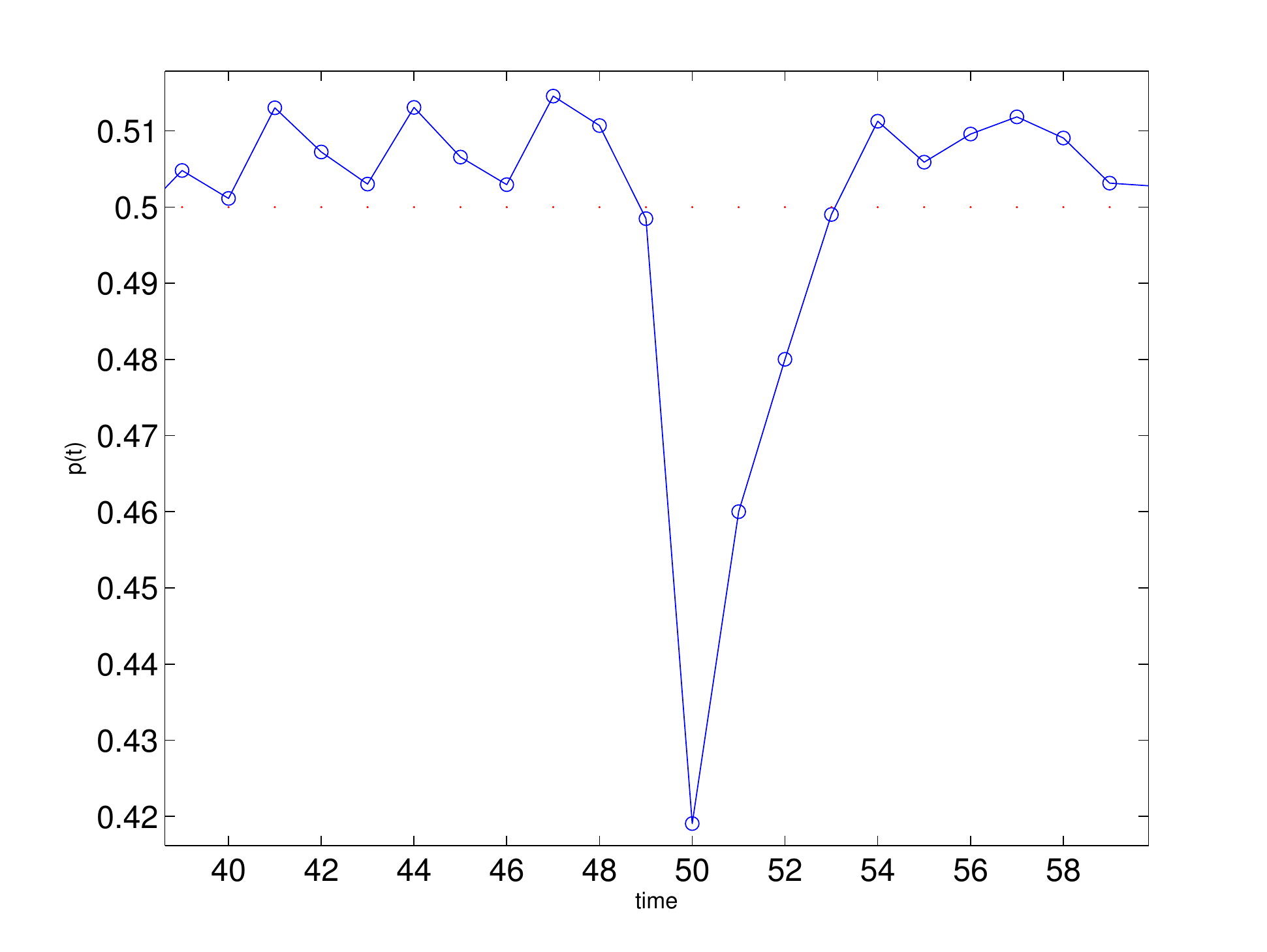}
                \caption{ }
                \label{fig:many_dropsc}
        \end{subfigure}
       
        \caption{\textbf{Temporary \textit{large} deviations from a fundamental can be explained by opinion shifts}. Opinion shift (drop) in $2$ essential classes leads to drop in price (blue line) and return to fundamental after confidence weights have shifted  due to the update to the fundamental (red dotted line). The figure on the right zooms in the time periods where the opinion shift takes place. This figure provides a possible interpretation of the DJ drop on April 13, 2013 after the fake tweet apperared.}\label{fig:many_drops}
\end{figure}

\bigskip
\bigskip

In order to understand the DJ30 drop occurred in the early afternoon of April 23, 2013 described in the Introduction of this work (and reported again in Fig. \ref{fig:many_dropsa}) we focused on a one-run simulation reported in Fig. \ref{fig:many_dropsb} and, zoomed-in, in \ref{fig:many_dropsc}. Here, we simulated a situation where drops in opinion occurred in only \textit{two} essential classes for a total number of $10$ agents shifting opinion. The drop is dramatic and shows the importance of the topological classification of agents.

\section{Empirical validation and model calibration}
\label{sec:empiricalvalidation}

The empirical validation of such a model is not an easy task. Although some stylized facts like clustered volatility, drops in prices that lead to fat tails in the returns distribution can be explained by simulations, the calibration of the parameters of the model we have is an open problem that we leave for future research. For example, as we have seen, the correct estimation of the update propensity parameter $\alpha$ is crucial.  

Time series for prices are easily available. However, determining and calibrating the interaction network and the update propensity parameter seems to be the hardest part of the problem. We propose to use the wide availability of \textit{Big Data} (i.e. Twitter microblogging) to have a proxy for such interactions. Applying Twitter data (and, in general, social network big data) in the context of financial markets is still a research endeavour (see Bollen, 2010). However, the recent use of Twitter data in giving signals about markets by the famous data service provider Bloomberg, clearly shows a growing interest in this approach. We can then give an outline of the future research that will need to be done to accomplish this task:

\begin{itemize}
\item Twitter and, more generally, social network provide with a huge amount of data, in the order of terabytes per minute. As such, algorithms must be fast and efficient, especially in the classification of agents.
\item At each $t$, agents needs to be classified according to the their topological importance (this includes, but it is not limited to, the separation into essential and inessential classes).
\item Social network data naturally embed a \textit{multiple network structure}, i.e. a structure where several interactions might occur. A possible way to cope with this problem is to use a tensor sequence as a means to treat multiple networks (see Kolda, 2005) $\left(\tensor{A}(t)\right)_{t \in \mathbb{N}}$

\begin{equation*}
a_{ijk}(t) = \begin{cases}
\in (0, 1] \mbox{ if $j$ retweets $i$'s tweet about stock \$$k$ at $t$} \\
0 \mbox{ otherwise}
\end{cases}
\end{equation*}

\end{itemize}



\section{Conclusions and future research}

In this work, we have proposed an analytical framework to model the role of opinion dynamics in a financial system. We have stressed the intertwined relationship among opinions, the actual price dynamics and the interaction network among agents. Different cases have been explored and we showed, by simulations, that some stylized facts can be reproduced and explained in this way. Moreover, we proposed an empirical validation scheme for the model.

This work stems from the need to interpret the role of opinion dynamics in an increasingly interconnected world, where network effects play an important role in determining the evolution of social systems.

Particular emphasis has been given to the topology of the interaction network and to the classification of agents according to their role in the dynamical process of opinion formation. We have explored the different models proposed in the literature and we have built on a well -- known nonlinear model where opinion fragmentation might occur. In this case, we noticed that such dynamics reflects on the price formation process in a feedback loop in the sense that agents will tend to shape their opinion patterns according to the topology and vice -- versa, the topology is determined by the opinion patterns.

\clearpage

\subsection*{Future research directions}

The approach proposed in this work provides a framework that can be extended by future research on these topics. 

For what concerns the network interaction topology, a detailed analysis on the properties of network dynamics when the network has a specific topology should be done. We refer to, e.g. random, small -- world, scale -- free or core -- periphery (see Newman, 2010 for details on such network topologies) topology structures. Finding analytical results for specific topologies would be useful for empirical comparison and analysis of the dynamical system.  In particular, we think of a scheme where the topology is imposed to the essential classes: this would drive different polarisation/fragmentation patterns that would then need to be compared with the actual empirical data.

\begin{figure}[!htbp]
\includegraphics[scale=0.5]{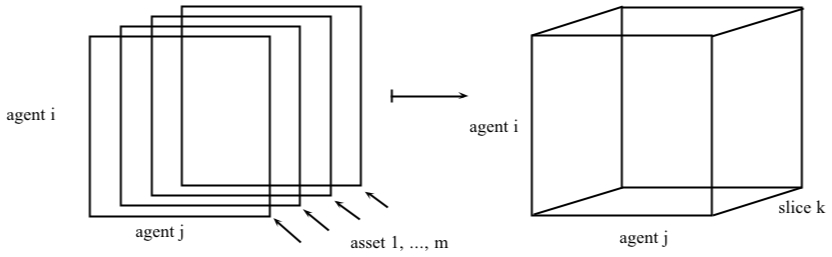}
\caption{Tensor multiple network approach for opinion dynamics.}
\label{fig:multiple}
\end{figure}

Moreover, we would like to extend the model on price formation to more than one risky asset (see Chiarella \textit{et al.} 2007 for an account on this problem) by assuming different confidence  networks for each asset. In other words, we want to investigate a \textit{multidimensional opinion dynamics} problem. We propose a dynamic \textit{multiple network} approach to this problem, by using a \textit{tensor} representation (see Kolda and Bader 2009), as follows (see Fig. \ref{fig:multiple} for a graphical representation):

\begin{displaymath}
\tensor{A}(t) \in \mathbb{R}^{n \times n \times m} = [a_{ijs}(t) ]
\end{displaymath}

\clearpage
\appendix

\section{Materials and methods}

\subsection{Graphs and networks}

We will outline the key concepts in graph and network theory used in the paper\footnote{The reader may refer to Harary (1969) and Newman (2010) for more details. Two relevant references for matrix theory are Gantmacher (1959) and Horn \& Johnson (1990).}.
A graph $G = (V, E)$ is a pair of sets $(V, E)$, where $V$ is the set of $n$ vertices and $E$ is the set of $m$ pairs of vertices of $V$. The pair $(i, j)$ belonging to $E$ is called an \textit{edge} of $G$ and $i$ and $j$ are called \textit{adjacent}. A \textit{directed graph} (digraph) is a graph in which each edge is an ordered pair $(i, j)$ of vertices. $G$ is simple if there one edge between two adjacent vertices. A weight $w_{ij}$ can be associated to each edge $(i, j)$ and, in this case, we will have a \textit{weighted} or \textit{valued} graph. A \textit{path} is a sequence of distinct adjacent vertices and a $i - j$ path is a path from $i$ to $j$. For a pair of vertices $(i, j)$, $i$ and $j$ \textit{communicate} if it holds $i- j$ and $j - i$. In case $w_{ii} >0$ $\forall i$, it always holds $i - i$, and we call $i$ \textit{self communicating}. A vertex $i$ is  \textit{essential} if, $\forall j \in V$, with $i - j$, it holds $j - i$ and we denote such property with $i \sim j$. A vertex $i$ is  \textit{inessential} if it is not essential.

$V$ can be partitioned into self--communicating equivalence classes of vertices. In fact it easy to see that if $i$ is self communicating and $(i, j)$ communicate, then $i \sim i$, $i \sim j$ $\Rightarrow$ $j \sim i$ and, finally, if $i \sim j$ and $j \sim k$ $\Rightarrow$ $i \sim k$, i.e. a path from $i$ to $k$ and vice versa can always be found passing through $j$. Hence, $\sim$ is an equivalence relation. A self-communicating inessential vertex constitutes an equivalence class on its own. In an essential class, all vertices communicate to each other and do not communicate with other vertices belonging to other classes. On the other hand, inessential vertices are linked to either essential and inessential vertices.\footnote{This terminology is the one used in Lorenz \cite{lorenz2006c}. In Markov chain theory, essential classes would be associated to a \textit{recurrent} class of communicating states and inessential vertices would be associated to \textit{transient} states.} A particular case arises when all vertices belong to one essential class, i.e. $\forall (i, j)$ $i \sim j$. In this case, the graph is called \textit{strongly connected}.

A nonnegative $n$ -- square matrix $\matr{W}$ representing the adjacency relationships (in general, weighted and directed), between vertices of $G$ is said to be the \textit{adjacency matrix} of $G$. A nonnegative adjacency matrix $\matr{W}$ is said to be \textit{diagonal dominant} if $w_{ii} \geq \sum_{j \neq i} w_{ij}$ $\forall i$. $\matr{W}$ is said to be \textit{irreducible} if for some permutation matrix $\matr{P}$, the (permuted) matrix $\mathbf{P} \mathbf{W} \mathbf{P}^T$ is \textit{not} block upper triangular. A matrix that is not irreducible is said to be \textit{reducible}. The adjacency matrix of a strongly connected graph is irreducible. If there exist a $t \in \mathbb{N}$ such that $\matr{W}^t$ is (strictly) positive, then $\matr{W}$ is said to be \textit{primitive}. An $n$--vector $\matr{x}$ is said to be a \textit{right eigenvector} of $\matr{W}$, with associated \textit{eigenvalue} $\lambda$ if $\matr{W}\matr{x} = \lambda \matr{x}$. For any $n \times n$ matrix, there exist $n$ eigenvalues $\lambda_1, \lambda_1, \ldots, \lambda_n$.

A nonnegative $n$--square matrix $\matr{W} \in \mathbb{R}^{n \times n}$ is row -- stochastic if the elements in each row sum up to $1$, i.e. $\sum_{j = 1}^n w_{ij} = 1 \;\; \forall i$ or, in matrix notation, $\matr{W}\matr{u} = \matr{u}$, where $\matr{u}$ is the $n$--th unit vector. It is easy to show that a row stochastic matrix has dominant eigenvalue $\lambda_1 = 1$ with $\matr{u}$ its associated eigenvector.

Any square matrix $\matr{W}$ can be brought in its \textit{Gantmacher form}:

\begin{displaymath}
\matr{W}_G = \matr{P} \matr{W} \matr{P}^\top = \left[
\begin{array}{cccccc}
\matr{W}_{1, 1} & & & & & \matr{0} \\
	& \ddots & & & & \\
	\matr{0} & & \matr{W}_{g, g} & & & \\ 
	\matr{W}_{g+1, 1}& \ldots & \matr{W}_{g+1, g}& \matr{W}_{g+1, g+1}&& \\
\vdots & &\vdots & \vdots & \ddots& \\
\matr{W}_{p, 1}& \ldots & \matr{W}_{p, g}& \matr{W}_{p, g+1}& \ldots& \matr{W}_{p, p}\\
\end{array}\right]
\end{displaymath}

where $\matr{P}$ is a suitable permutation matrix. The diagonal blocks $\matr{W}_{1, 1}, \ldots, \matr{W}_{p, p}$ are square and irreducible (Gantmacher 1959,  Lorenz 2007). For the non diagonal Gantmacher blocks $\matr{W}_{kl}$ with $ k = g + 1, \ldots, p$ and $l = 1, \ldots, k-1$. Then, it holds that for every $k = g+1, \ldots, p$ at least one block of $\matr{W}_{k, 1}, \ldots, \matr{W}_{k, k-1}$ contains at least one positive entry. It turns that $\matr{W}_{kk}$, with $k = 1, \ldots, g$ are the sub matrices associated to essential classes. Two non -- negative matrices $\matr{A}$ and $\matr{B}$ are said to be of the same \textit{type}, $\matr{A} \sim \matr{B}$ if they share the same non negative pattern, i.e. $a_{ij} >0 \Leftrightarrow b_{ij} >0$.

An important results of graph theory (see Harary, 1969) is that a non-connected graph $G$ can be \textit{uniquely} partitioned into separate strongly connected components (SCC).
When each one of these strongly connected component is contracted to a single vertex, the resulting graph is a directed acyclic graph (DAG), the so called \textit{condensation} of $G$.  A DAG is a digraph with no directed cycles, i.e. there exists no path connecting a node $i$ to itself. 

Given a graph $G = (V, E)$, its condensation digraph $C(C) = (V_C, E_C)$ is a graph where the set of nodes $V_C$ is represented by the strongly connected components of $G$. Accordingly, the condensation digraph of $G$, denoted $C(G)$, is defined as follows: the nodes of $C(G)$ are the SCC’s of G, and there exists a directed edge in $C(G)$ from node $H_1$ to node $H_2$ if and only if there exists a directed edge in $G$ from a node of $H_1$ to a node of $H_2$.

\subsubsection{Determining the Gantmacher form \label{find_gantmacher}}

We hereby propose a simple\footnote{This algorithm is inspired by Fig. 2 in Mirtabadei \& Bullo (2011). We thank Jan Lorenz (private communication) for providing useful hints.} in order to solve the problem. The algorithm outline is as follows:

\begin{enumerate}
\item find the Strongly Connected Components of $G$, e.g. via Tarjan's algorithm, of Depth First Search based algorithms ($O(|V| + |E|)$), see Knuth (1997);
\item find the condensation digraph $C(G)$ of $G$;
\item for each node $k$ of $C(G)$ check links for nodes;
\item if node $k$ has only in--coming links (i.e. sinks), then it is associated to an essential class;
\item if node $k$ has only out -- going links, then is is associated to an innessential class.
\item order nodes in $C(G)$ in order to find permutation matrix for $\matr{A}$.
\end{enumerate}

The algorithm proves to be quite fast for the applications in Section \ref{sec:empiricalvalidation}. However, better solutions that do not imply the evaluation of a node of $G$ multiple times could be found (via a Depth-First Search approach.). 

On the basis of such classification, a further refinement of Lorenz's classification has been proposed by Mirtabadei and Bullo (2011). See Table \ref{T:bullo} for a comparison of the two classifications.

\begin{table}
\caption{Agents classification (via SCCs)}
\centering
\begin{tabular}{cccc} \hline
 Lorenz (2006c)  & Mirt. Bullo (2011)  & SCCs condensation & SCC Subgraph \\ \hline
     essential & closed minded & sink & complete \\
  & moderate minded & not a sink & non -- complete\\ 
 inessential &  open minded & not a sink & either \\
 \hline
 \label{T:bullo}
\end{tabular}
\end{table}

\smallskip

We now deal with the mathematical background related to the opinion dynamics scheme. Several concepts are key in this class of models. We refer, in particular, to how accurate are agents' opinions about reality and how opinion pooling plays a role in this context\footnote{The so-called ``Wisdom of the Crowd'' idea, tracing back to Galton (1907).}. However, for sake of brevity, will shall not  explore these concepts in detail.

Consider a  set of \textit{interacting} agents and a certain dynamical process leading to the formation of an opinion space, represented by a scalar value. Some interesting questions naturally arise. Will all agents reach the same opinion? Will there be a consolidation of different opinions into fewer ones? We will refer to the first case as \textit{consensus} of opinions and to the latter as \textit{fragmentation} of opinions (see, for example, Hegselmann \& Krause, 2005).

DeGroot (1974) proposed one of the first mathematical formulations for the dynamics of opinions, by introducing a weighted averaging scheme that will remain the baseline model in later works. The convex weighted average scheme clearly stems from the mathematical properties related to the product of stochastic matrices and, hence, Markov chains.\footnote{DeGroot clearly refers to an opinion as a \textit{subjective probability distribution}, which each agent assigns to a certain parameter. Such \textit{a priori} opinions are then ``pooled'', to obtain an \textit{a posteriori} set of opinions, thus implying a Bayesian framework.  Previous formulations of this problem include the ones by J. French (1956), Harary (1959) and later contributions include the ones (among others) by Chatterjee and Seneta (1977) and Cohen \textit{et al}. (1986). These works focus on reaching a common opinion, a \textit{consensus}.}

In this light, another interesting work is the one of Berger (1981), in which sufficient conditions for consensus are discussed in more details and an \textit{implicit} relationships, later analyzed in the literature on economic networks (Jackson, 2008), between \textit{consensus} and eigenvector centrality is provided. Berger's paper shows also how a common opinion might be achieved only within sub--groups of agents, hence implying a certain degree of opinion fragmentation within the opinion structure(i.e. \textit{dissent} as opposed to \textit{consensus}). Other interesting approaches linking network centrality and opinion dynamics are those of Friedkin (1986, 1991) and Friedkin \& Johnsen (1990).

The literature so far reviewed is based on models of interaction that are \textit{linear} in their nature. In the models proposed by  DeGroot and Friedkin's, for example, the focus is on a linear dynamical system. As already mentioned, they make use of very well known properties of linear systems and, in particular, of homogeneous Markov chain. Berger's work implicitly introduces a graph -- theoretical argument based on connectivity, and Friedkin's approach is entirely graph -- based. 

A limitation lies in that such models do not capture the evolution in time of the interaction network. Concerning this problem, interesting analytical results for time--dependent networks are found in Moreau (2005) in the context of synchronization analysis and by Fazeli \& Jadbabaie (2012), who introduce a martingale network process associated to a Polya Urn process. 

So far, opinions and network topology are separated. However, in our model of price formation, there is a strong interaction between opinions and the confidence network. In other words: the interaction topology might be influenced by the opinions themselves.

An interesting way to cope with both these problems is the approach proposed by Krause in a series of works (1997, 2000 and, with Hegselmann, in 2005). In this model, the non--linearity arises because the topology of the confidence network changes in time according to the opinion structure and, viceversa, the new opinion structure is derived by the network interaction. This feedback relationship does not allow to easily derive analytical results. A combination of matrix and graph -- theoretical arguments can be used in order to partially overcome this problem, but most of the interpretation must rely on computer simulations. In fact, although it is possible to derive some sufficient conditions for the convergence of a system, it is still an open challenge to fully characterise in an analytical way the dynamical behaviour of such a system.

Hegselmann \& Krause (2005) propose a \textit{bounded confidence}  non -- linear approach: agents put a positive confidence weight to another agent if the absolute value of the difference of the opinions does not exceed a certain threshold. According to such a threshold, this non -- linear model can capture effects of opinion fragmentation or polarization as well as convergence to a consensus.

We define a stable opinion pattern in the dynamical system as that situation where agents stabilize their opinion and no further change occurs. Once such a stable pattern is achieved, we can, in fact, have the two following situations:
\vspace{-2mm}

\begin{enumerate}
\item \textit{consensus}, i.e. all agents have the same opinion;
\item \textit{dissent}, i.e. agents will have different opinions that do not change over time. In particular, we can divide \textit{dissent} into \textit{polarization} (i.e. agents can have only two opinions) and \textit{fragmentation} (when the number of possible opinions is higher than two).
\end{enumerate} 
\vspace{-2mm}
Sufficient conditions for convergence in these models are found in the above mentioned works. In addition to this, we recall the work of Lorenz (2005, 2006, 2007). We will observe, in particular, that the setup suggested by Lorenz provides with useful insights on the topological classification of agents. Further insights and some more technical results on the stabilisation patterns in the classification of agents can be found in Mirtabadei and Bullo (2011).

\subsection{Opinion dynamics: the mathematical background}

\subsubsection{The homogeneous opinion dynamics process}

Consider a set $V$ of $n$ agents, and an \textit{opinion vector} $\matr{x}(t) \in \mathbb{R}^n$ at time $t \in [0, \ldots, +\infty)$. The element $x_i(t)$ represents the opinion of agent $i$ at time $t$. The $n$ agents can be seen as vertices on a network, discussing a particular issue or trying to obtain better guesses about the value of a variable at each time step, such as the price of a financial asset. 

The general \textit{homogeneous} consensus dynamic model can be written as (DeGroot, 1974):

\begin{equation}
\matr{x}(t+1) = \matr{A} \matr{x}(t)
\end{equation}

Where $\matr{A}$ is a row -- stochastic matrix, i.e. $\sum_j a_{ij} = 1$, $\forall i$. We will refer to this matrix as the \textit{confidence} matrix. The process is said to be \textit{homogeneus} as the confidence matrix does not change in time (time--invariant process). At time $t+1$, we have:

\begin{equation}
\label{classical_mod}
\matr{x}(t+1) = \matr{A}^{t+1} \matr{x}(0)
\end{equation}

The model in Equation \ref{classical_mod} reaches a \textit{consensus} if  $\forall \matr{x}(0) \in \mathbb{R}^n$, $\exists c$ s.t. $\lim_{t \rightarrow \infty} \matr{x}(t) = \matr{A}^{\infty} \matr{x}(0) = \matr{c}$, where $\matr{c}$ is a real -- valued vector with all elements equal. The following two--part theorem gives conditions in this case:\footnote{For sake of brevity, we do not report the proofs of these theorem. We refer the reader to the already cited works of Gantmacher (1959), DeGroot (1974), Berger (1981) and Hegselmann and Krause (2005).}

\begin{theorem}[Conditions for convergence to consensus]
A \textit{consensus} for system \ref{classical_mod} is achieved:
\begin{itemize}
\item  if $\forall (i, j) \in E$, $\exists k \in V$ such that $a_{ik} >0$ and $a_{jk} > 0$ [see DeGroot (1974) for the proof];
\item if and only if $\exists$ $t_0 \in T$ s.t. the matrix $\matr{A}^{t_0}$ contains \textit{at least} one strictly positive column [see Berger (1981) for the proof].
\end{itemize}
\end{theorem}

\begin{theorem}[Conditions for convergence]
Let $\matr{A}$ be in its Gantmacher form with diagonal Gantmacher blocks  $\matr{A}_k$, $k = 1, \ldots, s$ and $g = 0, \ldots s$, then 
\begin{itemize}
\item $\lim_{t \rightarrow \infty} \matr{x}(t)$ exists $\forall \matr{x}(0) \in \mathbb{R}^n$ \textit{if and only if} the Gantmacher blocks are all primitive;
\item the system reaches a consensus \textit{if and only if} $g = 1$ (i.e. there exists only one essential class and no inessential classes).
\end{itemize}
\end{theorem}

This first general theorem states that a sufficient condition for consensus is found when any two agents have a positive weight on a same third agent. This conditions is related to the primitivity of the matrix. The value $c$ of the consensus depends, clearly, on the initial opinion profile $\matr{x}(0)$.

The second theorem employs the Gantmacher form. If each of the sub -- groups of essential agents has a primitive structure, then a stable opinion configuration is achieved where only the opinions of the essential agents matter in determining the asymptotic opinion vector. More over, we have a necessary and sufficient condition for consensus when the sub -- group of essential agents is only one ($g = 1$, i.e. only one essential class).

We define a \textit{consensus matrix} $\tilde{\matr{A}}$ the rank -- one strictly positive row -- stochastic matrix whose rows are all equal. It is immediate to see that $\tilde{\matr{A}}\matr{y}$ always gives a consensus.

%
%
%

\subsubsection{Inhomogeneous opinion dynamics process}

A more general approach is obtained when $\matr{A}$ is time dependent:
 
 \begin{equation}
 \label{inhom_mod}
 \matr{x}(t) = \matr{A}(t) \cdots \matr{A}(0) \matr{x}(0) 
 \end{equation}

We refer to the process described by Eq. \ref{inhom_mod} as either \textit{inhomogenous} (using the Markov chains expression) or \textit{time -- variant} (as opposed to the \textit{time -- invariant} process described in Eq. \ref{classical_mod}). Quite clearly, the ``inhomogenous'' or ``time -- variant'' part of the model stems from changes  in the agents' interaction topology and not directly from the opinion patterns in time. However, we will see that, in the case when the weight (and hence the interaction topology) depends on the opinions themselves, the feedback relationships between the opinions and the network topology will show non -- trivial behaviours.

The definitions we previously gave in the homogeneous case naturally extend to the inhomogeneous case. The sequence of $n \times n$ non -- negative matrices $\mathbf{A}(t)$ in our model is said to be a sequence of \textit{confidence} matrices if and only if the matrices:
\begin{enumerate}
\item are row -- stochastic, i.e. $\sum_{j} a_{ij}(t) = 1$, $\forall t$;
\item have strictly positive diagonal values, i.e. $a_{ii}(t) >0$, $\forall i \in V$ and $\forall t$;
\end{enumerate}

The objective of our analysis will be thus the sequence of matrices $(\matr{A}_0, \ldots \matr{A}_t)$ and, in particular, their right -- product $\matr{A}(t) \ldots \matr{A}(0)$ and its asymptotic behaviour for $t \rightarrow \infty$.\footnote{As opposed to the Markov chain process, in which case, we are interested in the left product of the elements of the sequence: $\matr{A}(0) \ldots \matr{A}(t)$.}

The process described by Equation \ref{inhom_mod} is clearly more difficult to treat than the homogeneous case and analytical results are harder to obtain. As usual, we are interested in determining sufficient and necessary conditions for the convergence of the system.  The structure of the sequence of matrices $\matr{A}(t)$ will determine the asymptotic behaviour of the system. We will observe that, analogously to what happens in the homogenous case, consensus or, more generally, convergence to consensus in the sub -- groups will be obtained as long as the weights keep being positive to a certain degree and/or sufficient time passes by.

Several interesting results have been proposed in the literature. Among these, we recall Moreau (2005), who describes a model of dynamic network interaction based on synchronization theory, providing results based on both graph theory and system theory; Fazeli and Ali Jadbabaie (2010), who use a Polya urn argument to show certain asymptotic properties and Fazeli \& Jadbabaie (2012), who propose some analytical results on convergence where the underlying network process is a martingale. Alternative averaging schemes are not investigated in this work.\footnote{For a discussion on alternative averaging schemes see Krause (2000 and 2008). For a detailed explanation on the speed of convergence of the system, refer to Isaacson \& Madsen (1976),  Gross \& Rothblum (993); Rothblum \& Tan (1995). See Dobrushin, (1956) for an analysis of the ergodicity coefficient, later tackled by Ipsen \& Selee (2011) in an analysis that also explores concepts in network centrality.}

\subsubsection{Convergence in the inhomogenous case}

The intution behind the convergence in the inhomogeneous case is that a time -- variant model will converge provided that, as previously noted, both the weights keep beeing sufficiently positive and/or that enough time passes. The positive diagonal assumption will also play a significant role in ensuring this. Clearly, should the weights go to zero too fast, convergence (let alone consensus) might not be achieved. 

Krause (2000) proposes an interesting theorem for convergence (not limited to arithmetic means) and later, Hegselmann \& Krause provide more general approach based on the arithmetic average setting  based on the \textit{accumulated weights}. The intuition behind this approach is to keep track of  the weights of the chain of products.

Another way to see this problem is to exploit the positive diagonal. In fact, given a non -- negative matrix $\matr{B}$ with positive diagonal and a non -- negative matrix $\matr{C}$, then the product $\matr{B}\matr{C}$ has at least the same positive entries of $\matr{C}$. Based on the above reported results, Lorenz (2005 and 2007) reports a  convergence theorem based on the Gantmacher form.

\begin{theorem}
Given $\matr{A}(t)$, $t = 0, 1, 2, \ldots, $ non -- negative, row -- stochastic with strictly positive diagonal, and $\min^+ \matr{A}(t + 1)  \matr{A}(t) \geq \delta_t$ and $\sum_{t = 1}^\infty \delta_t= \infty$ Gantmacher blocks converge to:

\begin{displaymath}
\lim_{t \rightarrow \infty} \left(\matr{A}(t) \cdots \matr{A}(0)\right) = \left[
\begin{array}{ccc|c}
\matr{G}_1 & & & \matr{0}  \\
	& \ddots & &  \\
	\matr{0} & & \matr{G}_g & \matr{0}\\ \hline
\mbox{n.c}& \ldots & \mbox{n.c.}& \matr{0}\\
\end{array}\right]
\end{displaymath}

where $g$ is the number of essential classes, $n_h$ is the cardinality of essential class $h$ and $\matr{G}_h$ are $n_h \times n_h$ row -- stochastic consensus matrices, i.e. strictly positive stochastic matrices with all rows equal:

\begin{displaymath}
\matr{G}_h = \left[
\begin{array}{cccc}
c_1 & c_2 & \ldots & c_n \\
\vdots & \vdots & \vdots &\vdots \\
c_1 & c_2 & \ldots & c_n
\end{array} \right] = \left[\begin{array}{c|c|c|c}
& & & \\
\matr{c}_1& \matr{c}_2& \ldots &\matr{c}_n \\
& & & \\

\end{array}\right]
\end{displaymath}

\end{theorem}

\subsubsection{Opinion dynamics under bounded confidence}

The inhomogenous case is often referred to as a \textit{time -- variant} model, in the sense that the weights each agent puts onto other agents changes over time. However, a rule for the evolution of the network topology must be established in order to analyze the system dynamics.

The main question in this case is whether the update rule will influence the reaching of a stable opinion pattern or fragmentation (either consensus within subgroups or dissent across different subgroups) will occur or even no convergence at all will be achieved.

Within this framework, ``the most difficult type of model occurs if the weights depends on opinions itself because then the model turns from a linear one to a \textit{non--linear} one'' (cit. Hegselmann \& Krause, 2005).

In particular, we hereby examine a system with \textit{bounded confidence}, in which agents update their network interaction structure by putting weight on agents that have a similar opinion. The system becomes \textit{non -- linear} in the sense that the  averaging process can automatically include or discard certain agents, hence underweighting certain opinions or overweighting other opinions if they are similar. 

We define a \textit{bounded confidence} opinion dynamics process, as the system with initial opinion vector $\matr{x}(0)$ and dynamics:

\begin{displaymath} 
\matr{x}(t+1)  =  \matr{A}(t+1)  \matr{x}(t) = \matr{A}(x(t), \matr{\epsilon)} \matr{x}(t)
\end{displaymath}

\begin{equation}
\label{eq:boundedconfidence}
a_{ij}(t+1) := \left\{\begin{array}{cc}
\frac{1}{\# I(i, \matr{x}(t))} & \mbox{if } j \in I(i, \matr{x}(t)) \\
0 & \mbox{otherwise}
\end{array} \right.
\end{equation}

\begin{displaymath}
I(i, \matr{x}(t)) = \{j \mbox{ s.t. } |x_i(t) - x_j(t)| \leq \epsilon_i\}
\end{displaymath} 

This structure implies that, at each time $t$, every agent compares her opinion with that of others, finding a subset of agents whose opinion does not differ too much from her. The agent then assigns \textit{equal} weight to these agents. The assumption of equal weights can be relaxed, but it will not be analyzed in this work. More importantly, agent $i$ herself lies in the set $I$, thus leading to a positive diagonal at each time $t$.

The non -- linearity in the model is due to the fact that the update rule enhances the selection of similar opinions and implies discarding distant opinions. This approach naturally brings to  \textit{fragmentation} and \textit{polarisation} of opinions, meant as a \textit{hardening} of opinion patterns. Hegselmann \& Krause  (2005) describe, by simple examples, the possibility of splits between subgroups that become essential and that reach consensus within the agents belonging to the subgroup.

\paragraph{Simulations}

The models we have so far described are multi -- faceted and computer simulations can give  better insights. Since the homogeneous case is of little interest in this work, and it has been reported for sake of completeness, we refer to the non -- linear bounded confidence inhomogenous model in Equation \ref{eq:boundedconfidence} in the following simulations.

\begin{figure}[!htbp]
\centering 
\includegraphics[scale=1]{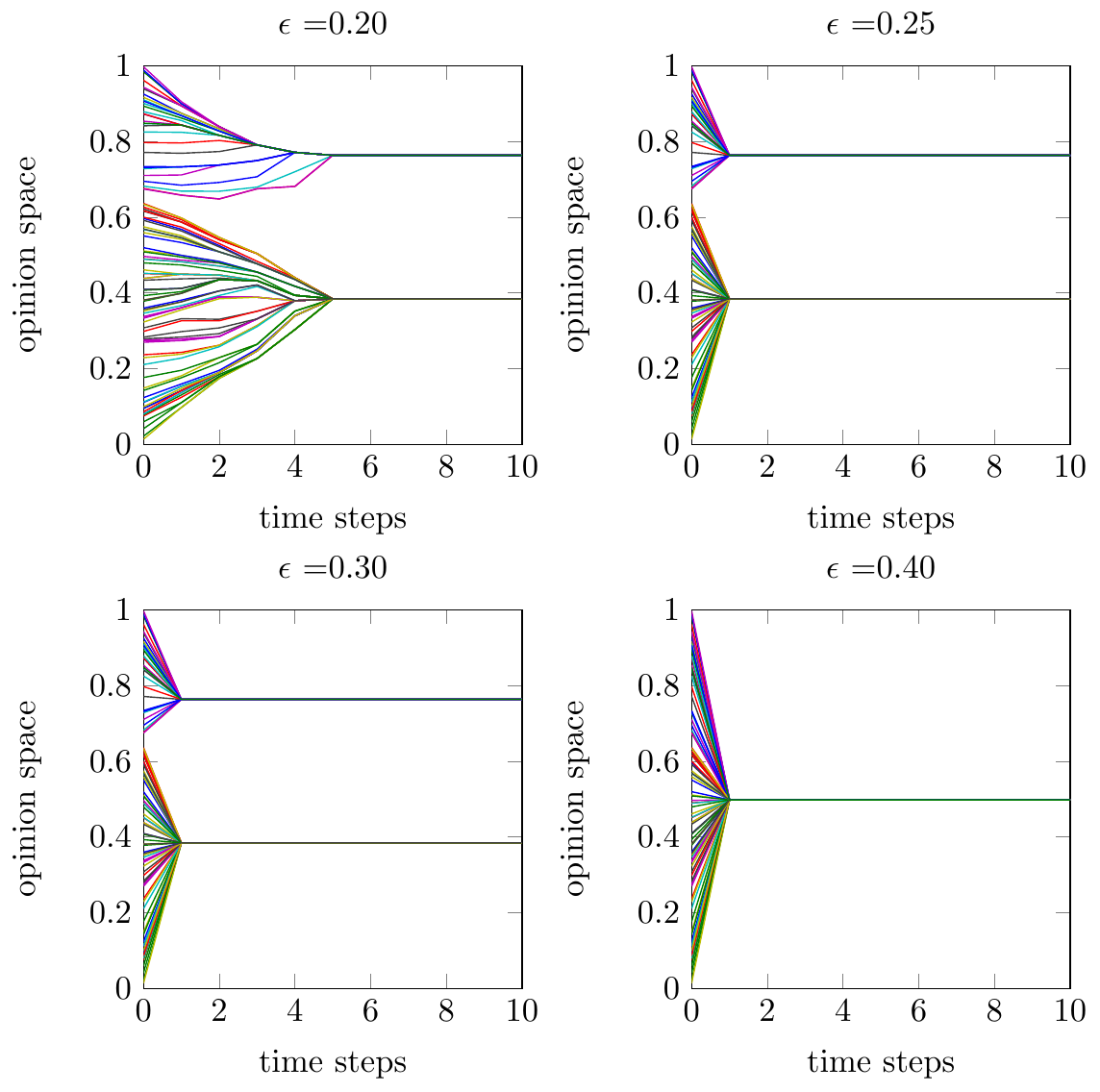} 
\caption{Single - run simulations} \label{fig:wrapsinglerun1} 
\end{figure}

All simulations are for $n = 100$ agents. The number of simulations, unless otherwise stated, is $1000$. In the first part, the initial opinion vector $\matr{x}(0)$ is drawn from a uniform continous random distribution with support $[0, 1]$. Figure \ref{fig:wrapsinglerun1} reports a simple one -- run simulation with different values of $\epsilon$ (assumed to be equal for all agents). It is immediate to see that the higher the $\epsilon$, the higher the chances to achieve consensus. However, polarized situations (e.g. $\epsilon = 0.20$) can be obtained, which show ``hardening'' in the opinions. When $\epsilon = 0.40$ a consensus is achieved in a very limited number of time steps.

\begin{figure}[!htbp] 
\centering 
\includegraphics[scale = 1.]{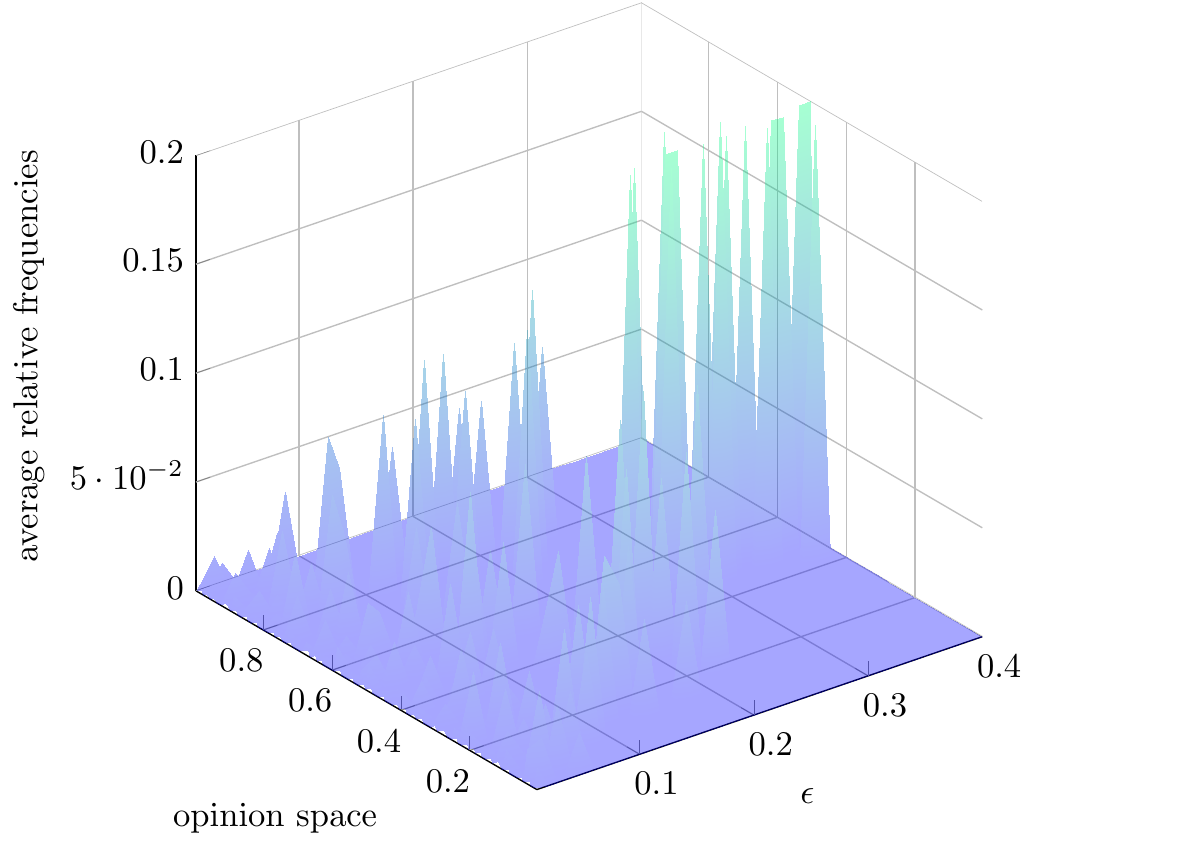} 
\caption{Average relative frequencies of $x \in [0, 1]$, $n = 100$ opinions (after stabilisation) for different values of $\epsilon$ ($1000$ simulations).} \label{fig:diag100} 
\end{figure}

Figure \ref{fig:diag100} shows a tri -- dimensional plot of the opinion space with respect to different values of $\epsilon$ \textit{once a stable configuration is achieved}, the $y$ axis is the average relative frequence over $1000$ simulations. One can see how the higher the $\epsilon$, the less fragmented the opinion pattern is.

%



\newpage

%

\paragraph{Distribution of the initial opinion profile}
Since in the non -- linear bounded confidence model the initial opinion profile can determine the degree of fragmentation of the opinion patterns, it is interesting to conclude this Chapter by modifying the distribution of the initial opinion profile $\matr{x}(0)$. Quite obviously, if the initial opinion profile is too dispersed (with respect to the confidence level $\epsilon$), the initial fragmentation pattern will never converge to a consensus or to a less fragmented pattern.

While still keeping the opinion range $\matr{x}(0) \in [0, 1]$, we will use an initial opinion profile from random numbers drawn from a Beta distribution\footnote{The pdf of a Beta distribution is given by \begin{displaymath}f(x; a, b) = \frac{1}{B(a, b)} x^{a-1} (1-x)^{b-1},\end{displaymath} where $B(a, b)$ is the Beta function and $a$ and $b$ are the shape parameters of the distribution.}.


%
%
%


\subsection{Software and data}

Data processing and simulations have been done in Python and Matlab. Tri -- dimensional plots have been exported to \LaTeX via \verb|matlab2tikz|.\footnote{ \url{http://www.mathworks.de/matlabcentral/fileexchange/22022-matlab2tikz/all_files}}

Twitter (\url{www.twitter.com}) data have been retrieved and pre -- processed (string and hashtag recognition) in Python with Python Twitter\footnote{The current analysis, as of May 2014, is done in Twython.} (a Python wrapper around the Twitter API (\url{https://dev.twitter.com/}) and double--checked via Twitty for Matlab. 

Network visualizations are done with Gephi (\url{https://gephi.org/‎}).


\clearpage

\setstretch{0.2}

\end{document}